\documentclass[useAMS,usenatbib]{mn2e}
\usepackage{graphicx}
\usepackage{amssymb}
\usepackage{mathrsfs}
\usepackage{natbib}
\usepackage{overpic}

\usepackage{color}
\definecolor{rplt}{rgb}{0.75,0,0}
\definecolor{gplt}{rgb}{0,0.75,0}
\definecolor{bplt}{rgb}{0,0,0.75}

\def\totd{{\mathrm{d}}}

\def\sun{{\odot}}

\title[SASI- and Convection-Dominated Explosions]{Characterizing SASI- and 
Convection-Dominated Core-Collapse Supernova Explosions in Two Dimensions}
\author[Fern\'andez, M\"uller, Foglizzo, \& Janka]{Rodrigo Fern\'andez$^{1,2,3}$,
Bernhard M\"uller$^4$, Thierry Foglizzo$^5$, Hans-Thomas Janka$^4$\\
$^1$ Institute for Advanced Study, Princeton, NJ 08540, USA\\
$^2$ Department of Physics, University of California, Berkeley, CA 94720, USA\\
$^3$ Department of Astronomy \& Theoretical Astrophysics Center, University of California, Berkeley, CA 94720, USA\\
$^4$ Max-Planck-Institut f\"ur Astrophysik, Karl-Schwarzschild-Str. 1, D-85748 Garching, Germany\\
$^5$ Laboratoire AIM, CEA/DSM-CNRS-Universit\'e Paris Diderot, IRFU/Service d'Astrophysique, CEA-Saclay F-91191, France 
}

\begin{document}

\date{Submitted to MNRAS}
\pagerange{\pageref{firstpage}--\pageref{lastpage}} \pubyear{2002}
\maketitle
\label{firstpage}


\begin{abstract}
The success of the neutrino mechanism of 
core-collapse supernovae relies on the supporting action of two hydrodynamic instabilities:
neutrino-driven convection and the Standing Accretion Shock Instability (SASI).
Depending on the structure of the stellar progenitor,
each of these instabilities can dominate the evolution of the gain region prior to the onset of 
explosion, with implications for
the ensuing asymmetries. Here we examine the flow dynamics in the neighborhood
of explosion by means of parametric two-dimensional, time-dependent hydrodynamic simulations
for which the linear stability properties are well understood. We find that systems
for which the convection parameter $\chi$ is sub-critical (SASI-dominated) develop
explosions once large-scale, high-entropy bubbles are able to survive for several
SASI oscillation cycles. These long-lived structures are seeded by the SASI during shock 
expansions. Finite-amplitude initial perturbations do not alter this outcome
qualitatively, though they can lead to significant differences in explosion times. 
Supercritical systems (convection-dominated)
also explode by developing large-scale bubbles, though the formation of these
structures is due to buoyant activity. Non-exploding systems achieve a quasi-steady 
state in which the time-averaged flow adjusts itself to be convectively sub-critical. 
We characterize the turbulent flow using a spherical Fourier-Bessel decomposition, 
identifying the relevant scalings and connecting temporal and spatial components.
Finally, we verify the applicability of these principles on the general relativistic,
radiation-hydrodynamic simulations of M\"uller, Janka, \& Heger (2012), and discuss
implications for the three-dimensional case.
\end{abstract}


\section{Introduction}

In the neutrino mechanism of core-collapse supernovae, 
a small fraction of the energy emitted in neutrinos by 
the forming neutron star is deposited in a layer behind the
stalled accretion shock, powering its final expansion \citep{bethe85}. 
Extensive theoretical work over the last two decades has 
led to a consensus on the failure of this mechanism
in spherically symmetric systems, except for the very lightest
stellar progenitors (see, e.g., \citealt{janka2012a} for a recent review).

Successful neutrino-driven explosions require additional assistance
by non-spherical hydrodynamic instabilities that
increase the efficiency of neutrino energy deposition. This 
phenomenon has been observed in numerous two-dimensional
(e.g., \citealt{herant94,burrows95,janka96,mezzacappa98,scheck06,ohnishi06,
buras06b,burrows07,murphy08,ott08,marek09,suwa09,mueller2012,couch2013})
as well as three-dimensional
(e.g., \citealt{iwakami08,nordhaus10a,hanke2012,burrows2012,
emueller2012,takiwaki2012,ott2012b,couch2012,dolence2013,hanke2013})
core-collapse simulations of various levels of sophistication. In addition to assisting
the onset of explosion, these instabilities can contribute to
the generation of pulsar kicks \citep{scheck06,nordhaus10b,wongwathanarat2010}, 
the spin-up of the forming neutron star \citep{fryer07,blondin07a,blondin07b,F10} 
and the seeding of late-time asymmetries \citep{kifonidis06,hammer2010,wongwathanarat2012}.

The shock-neutrinosphere cavity is unstable to \emph{convection} 
driven by the energy deposition from neutrinos emitted in deeper layers (e.g., \citealt{bethe90}).
This process generates kinetic energy on spatial scales comparable to or smaller than the size of the
neutrino heating region. 
Work by \citet{BM03} and \citet{BM06} isolated a distinct, global oscillatory instability of the standing
accretion shock that operates independent of neutrino heating, the so-called 
Standing Accretion Shock Instability (SASI). The driving mechanism involves
an unstable cycle of advected and acoustic perturbations trapped within the
shock-neutrinosphere cavity (\citealt{F07,foglizzo09,guilet2012}). The most unstable modes
of the SASI reside on the largest spatial scales. Convection and the SASI
are easily distinguishable in the linear regime, but their effects 
become intertwined in the non-linear turbulent flow that follows the stalling of the 
bounce supernova shock (e.g., \citealt{scheck08}).

Recent three-dimensional studies of core-collapse supernova hydrodynamics
have found that large-scale oscillation modes of the shock attain 
smaller amplitudes than in two dimensions \citep{nordhaus10a,wongwathanarat2010,hanke2012,
takiwaki2012,murphy2012}. This has been interpreted as a consequence of the
different behavior of turbulence in two- and three-dimensions
\citep{hanke2012}, and has led to the suggestion that the SASI may play a secondary
role in the explosion mechanism, if it arises at all \citep{burrows2012,burrows2013}.
These models have largely focused on a small sample of stellar progenitors, however,
and in many cases do not include physical effects that are favorable for the
growth of the SASI \citep{janka2012b}.

\citet{mueller2012} followed the collapse and bounce of $8.1$ and $27M_\sun$ progenitors
using a two-dimensional, general relativistic hydrodynamic code
with energy dependent neutrino transport, finding that differences in progenitor
structure lead to very different paths to explosion. In particular, the $27M_\sun$ 
progenitor evolution is such that the SASI dominates the dynamics throughout
the pre-explosion phase. Three-dimensional simulations of the same progenitor, with similar neutrino 
treatment, display episodic SASI activity, though a successful explosion is not yet obtained \citep{hanke2013}. 
\citet{ott2012b} evolved the same $27M_\odot$
progenitor with a more approximate neutrino prescription and a higher level of numerical perturbations, 
initially finding smaller SASI amplitudes than \citet{mueller2012} and \citet{hanke2013}, though
later confirming SASI activity (C. Ott 2013, private communication).
The lack of the same level
of numerical perturbations in the \citet{mueller2012} models could mean that 
convection dominance instead of SASI dominance
is dependent not only on the progenitor structure, but also on the details of the initial
conditions.

It is the purpose of this paper to investigate some of these
issues involving the interplay of SASI and convection, and the
implications for successful explosions. 
In particular, we address the following questions:
(1) Is there a fundamental difference between the transition to explosion in SASI-
and convection-dominated models? 
(2) Can finite amplitude perturbations, 
generated in, e.g.,  multi-dimensional stellar progenitors 
(e.g., \citealt{arnett2011}), tilt the balance in favor of convection
in situations that would otherwise be SASI-dominated? 
(3) Does the SASI play any discernible role in convection-dominated systems?
(4) Are there systematic trends in models close to an explosion that shed
insight into the operation of each instability?

Our approach is experimental, employing hydrodynamic simulations that
model neutrino source terms, the equation of state, and gravity in a 
parametric way (e.g., \citealt{FT09b}). 
This setup has the advantage that its linear stability properties
are well understood \citep{FT09a}, allowing the development
of model sequences that probe different parameter regimes. 
Our experimental
approach to studying SASI and convection follows similar
works \citep{foglizzo06,ohnishi06,scheck08,FT09b,burrows2012}, to which 
we relate our findings. To connect with more
realistic models, we also test the generality of our analysis results on the
simulations of \citet{mueller2012}. 

The structure of the paper is the following. Section~\ref{s:methods} describes
the numerical models employed and introduces the Spherical Fourier-Bessel
decomposition. Section~\ref{s:results} presents results, separated by
exploding and quasi-steady state behavior. A summary and discussion 
follows in Section~\ref{s:summary}. Appendix~\ref{s:L0} addresses the
stability of the $\ell=0$ mode in the parametric setup, and 
Appendix~\ref{s:sfb_appendix} provides 
details about the spherical basis functions for the cases of Dirichlet and 
Neumann boundary conditions.

\section{Methods}
\label{s:methods}

\subsection{Parametric Hydrodynamic Simulations}
\label{s:parametric}

\subsubsection{Numerical Setup}
\label{s:setup}

The parametric, two-dimensional stalled supernova shock simulations employed for
the majority of the analysis follow the setup of \citet{FT09b,FT09a}. 
These models have been calibrated to the global linear
stability analysis of \citet{F07}. The linear analysis has
been extended to include the effects of parameterized nuclear dissociation 
and lightbulb neutrino heating \citep{FT09b}.

In our time-dependent models, the equations of mass, momentum, and energy conservation are solved in 
spherical polar coordinates ($r,\theta$), subject to the gravity from a 
point mass $M$ at the origin and parameterized neutrino heating and cooling:
\begin{eqnarray}
\frac{\partial \rho}{\partial t} + \nabla\cdot(\rho\mathbf{v}) & = & 0\\
\frac{\partial \mathbf{v}}{\partial t} + (\mathbf{v}\cdot \nabla)\mathbf{v}
				     &  = & -\frac{1}{\rho}\nabla p - \frac{GM}{r^2}\hat r\\
\frac{\totd e_{\rm int}}{\totd t} -\frac{1}{\rho}\frac{\totd p}{\totd t} & = & Q_\nu.
\end{eqnarray}
Here $\rho$, $\mathbf{v}$, $p$, and $e_{\rm int}$ are the fluid density, velocity,
pressure, and specific internal energy, respectively. The equation of state is that of 
an ideal gas with adiabatic index $\gamma$, 
i.e., $p = (\gamma-1)\rho e_{\rm int}$.
To connect with previous studies, the net neutrino source term is set
to 
\begin{equation}
\label{eq:neutrino_source_term}
Q_\nu = \left[ \frac{B}{r^2} - A p^{3/2}\right]
			       \,e^{-(s/s_{\rm min})^2}\,\Theta(\mathcal{M}_0-\mathcal{M}),
\end{equation}
where $s$ is the fluid entropy, $\mathcal{M}$ the Mach number, and $\Theta$ the
step function.
This functional form models heating as a lightbulb, with $B$ a normalization
constant proportional to the neutrino luminosity. The cooling function,
first introduced by \citet{BM06} and subsequently used by \citet{F07}, models
electron and positron capture in an optically thin environment $(\propto \rho T^6)$
assuming a radiation-dominated gas $(p\propto T^4)$. The exponential suppression at a 
low entropy $s_{\rm min}$ is introduced to prevent runaway cooling at the base of the flow, 
and the cutoff at high Mach number $\mathcal{M}_0=2$ is used to suppress heating and
cooling in the upstream flow \citep{FT09b}.

The initial condition consists of a steady-state spherical accretion shock 
at a radius $r_s$, below which the fluid settles subsonically onto a 
protoneutron star of radius $r_*$. Given a boundary condition at the shock,
the normalization of the cooling function $A$ is determined by demanding that
the radial velocity vanishes at $r=r_*$.
The upstream flow is supersonic and adiabatic, with zero Bernoulli parameter. 
The Mach number upstream of the shock is set to $\mathcal{M}_1 = 5$ at a radius $r_{\rm s0}$
equal to the shock radius obtained with zero heating ($B=0$). 
To connect with previous studies (e.g., \citealt{FT09b}), the adiabatic index is set to
$\gamma=4/3$, even though a more realistic flow would have this index varying within the range $1.4-1.6$.
A constant specific energy loss by nuclear dissociation $\varepsilon$ is allowed at
the shock, increasing the compression ratio \citep{thompson00}.
The solution is uniquely determined by specifying the ratio $r_*/r_{\rm s0}$, 
the nuclear dissociation parameter $\varepsilon$,
the upstream Mach number $\mathcal{M}_1$, and the heating rate $B$ 
(see \citealt{FT09b} for a sample of initial density profiles). In all
models, we set $r_*/r_{\rm s0} = 0.4$.

Throughout this paper, we adopt the initial shock radius without heating, $r_{\rm s0}$,
the free fall speed at this radius, $v_{\rm ff0}^2 = 2GM/r_{\rm s0}$, and the
upstream density $\rho_1$ as the basic system of units. Full-scale simulations
yield characteristic values $r_{\rm s0}\simeq 150$~km, $M\simeq 1.3M_\odot$, and
$\dot{M}\simeq 0.3M_\sun$~s$^{-1}$, with a resulting free-fall speed
$v_{\rm ff0} \simeq 5\times 10^9$~cm~s$^{-1}$, dynamical time
$t_{\rm ff0} = r_{\rm s0}/v_{\rm ff0} \simeq 3$~ms, and upstream
density $\rho_1 \simeq 10^8$~g~cm$^{-3}$. Setting the heating term
in equation~(\ref{eq:neutrino_source_term}) equal to the approximation from \citet{janka01}
commonly used in `lightbulb' heating studies (e.g., \citealt{murphy08,couch2013}), one
obtains a relation between $B$ and the electron neutrino luminosity,
\begin{equation}
\label{eq:B_dimensional}
B \simeq 0.009\, L_{\nu_e,52}\, T_{\nu,4}^2\, 
\left(\frac{r_{\rm s0}}{150\textrm{ km}}\right)^{1/2}\left(\frac{1.3M_\odot}{M} \right)^{3/2},
\end{equation}
where $L_{\nu_e,52}$ is the electron neutrino luminosity in units of $10^{52}$~erg~s$^{-1}$,
and $T_{\nu,4}$ is the neutrinospheric temperature in units of $4$~MeV.

The numerical models are evolved in FLASH3.2 \citep{dubey2009}, with the modifications
introduced in \citet{F12}.
The computational domain covers the radial range $r\in [0.4,7]r_{\rm s0}$, and
the full range of polar angles. The radial grid spacing is logarithmic, with 408
cells in radius ($\Delta r/r \simeq 0.7\%$). We use 300 angular cells equispaced in $\cos\theta$,
yielding constant volume elements at fixed radius ($\Delta \theta \simeq \Delta r/r$
on the equator). The boundary conditions are reflecting at the polar axis and at the surface of the
neutron star, and set to the upstream solution at the outer radial boundary. Accreted material accumulates
in the innermost $\sim $ two cells next to the inner boundary.

\begin{table*}
\centering
\begin{minipage}{16.5cm}
\caption{Models Evolved and Results\label{t:models}\label{t:results}.
Columns show model name, dissociation parameter, 
heating constant (eq.~[\ref{eq:neutrino_source_term}]), initial convection
parameter (eq.~[\ref{eq:convection_parameter}]), minimum amplitude
for convection (eq.~[\ref{eq:delta_crit}]) assuming a spherical bubble with $l_v = 2(r_s-r_g)/3$, initial gain and shock radius, 
type and amplitude of initial perturbation (\S\ref{s:models}), 
advection time and convection parameter of time-averaged flow (eq.~[\ref{eq:chi_mean_flow}]), 
ratio of time-averaged kinetic energy to time-averaged mass in the gain region (non-exploding), and time $t_{\rm exp}$ 
at which the shock hits the outer radial boundary (exploding). The unit system is defined in \S\ref{s:setup}.}
\begin{tabular}{lcccccccccccc}
{Model}&
{$\varepsilon$} &
{$B$} &
{$\chi_0$} &
{$(\Delta\rho/\rho)_{\rm c}$} &
{$r_g$} &
{$r_s$} &
{Pert.} &
{Ampl.} &
{$\bar{t}_{\rm adv}$} &
{$\bar{\chi}$} &
{$E_{\rm kin,g}/M_{\rm g}$} &
{$t_{\rm exp}$} \\
\noalign{\smallskip}
{ } & {($v_{\rm ff0}^2/2$)} & {($r_{\rm s0} v_{\rm ff0}^3$)} & {} & {} & \multicolumn{2}{c}{($r_{\rm s0}$)} & {} &
{}    & {($t_{\rm ff0}$)} & {} & {($10^{-2}\,v_{\rm ff0}^2$)} &  {($t_{\rm ff0}$)}\\
\hline
e0B00    & 0  & 0     & 0    & ...  & ...  & 1    & rand. $\delta \mathbf{v}/v_r$  & $10^{-3}$ & 8.9  & 0     & ...  & ...\\
e0B02    &    & 0.002 & 0.06 & 0.14 & 0.90 & 1.04 &                                &           & 9.5  & 0.69  & 3.7  & ...\\
e0B04    &    & 0.004 & 0.3  & 0.07 & 0.78 & 1.09 &                                &           & 10.2 & 0.64  & 2.9  & ...\\
e0B06    &    & 0.006 & 0.6  & 0.05 & 0.72 & 1.14 &                                &           & 11.3 & 0.81  & 2.7  & ...\\
e0B08    &    & 0.008 & 1.0  & 0.05 & 0.69 & 1.20 &                                &           & 13.1 & 1.06  & 2.7  & ...\\
e0B10    &    & 0.010 & 1.5  & 0.04 & 0.67 & 1.28 &                                &           & ...  & ...   & ...  & 336 \\
\noalign{\smallskip}                                 
p0B08L1  & 0  & 0.008 & 1.0  & 0.05 & 0.69 & 1.20 & $\ell=1$ shell                 & 0.1       & 12.5 & 1.2   & 2.3  & ... \\
p0B10L1  &    & 0.010 & 1.5  & 0.04 & 0.67 & 1.28 &                                &           & ...  & ...   & ...  & 218 \\
p0B10L2  &    &       &      &      &      &      & $\ell=2$ shell                 &           & ...  & ...   & ...  & 376 \\
p0B10R1  &    &       &      &      &      &      & rand. $\delta\rho/\rho$        & 0.1       & ...  & ...   & ...  & 127 \\
p0B10R3  &    &       &      &      &      &      &                                & 0.3       & ...  & ...   & ...  & 241 \\
p0B10G4  &    &       &      &      &      &      & $\ell=4$ gain                  & 0.5       & ...  & ...   & ...  & 246 \\
p0B10G5  &    &       &      &      &      &      & $\ell=5$ gain                  &           & ...  & ...   & ...  & 122 \\
\noalign{\smallskip}
e3B00 & 0.3 & 0     & 0     & ...   & ...  & 1    & rand. $\delta \mathbf{v}/v_r$  & $10^{-3}$ & 19.8 & 0    & ...  & ... \\
e3B02 &     & 0.002 & 1.5   & 0.021 & 0.66 & 1.06 &                                &           & 22.5 & 0.89 & 0.9  & ... \\
e3B04 &     & 0.004 & 3.9   & 0.016 & 0.60 & 1.13 &                                &           & 25.5 & 1.33 & 1.9  & ... \\
e3B06 &     & 0.006 & 7.1   & 0.013 & 0.58 & 1.23 &                                &           & 34.8 & 2.11 & 2.6  & ... \\
e3B08 &     & 0.008 & 8.0   & 0.010 & 0.57 & 1.25 &                                &           & ...  & ...  & ... & 223 \\
\hline
\end{tabular}
\end{minipage}
\end{table*}

\subsubsection{Models Evolved and Initial Perturbations}
\label{s:models}

Based on the linear stability analysis of \citet{foglizzo06}, the transition from SASI- to 
convection-dominated behavior occurs when the parameter
\begin{equation}
\label{eq:convection_parameter}
\chi =  \int_{r_{\rm g}}^{r\rm s}\,\frac{\textrm{Im}(\omega_{\rm BV})}{|v_r|}\,\totd r,
\end{equation}
exceeds a critical value of the order of 3. Here $r_g$ is the gain radius, 
$\omega_{\rm BV}$ is the buoyancy frequency, 
\begin{equation}
\label{eq:brunt}
\omega^2_{\rm BV} = \frac{GM}{r^2}\left[\frac{1}{\gamma}\frac{\partial \ln p}{\partial r} 
		  -\frac{\partial\ln\rho}{\partial r}\right]
\end{equation}
and $v_r$ is the radial velocity.
This critical value of $\chi$ is the number of e-foldings by which an 
infinitesimal buoyant perturbation needs to grow 
to counter advection out of the gain region. Larger heating rates
and longer advection times are favorable for the growth of convection,
as they increase $\chi$. 

A finite-amplitude density perturbation can also overcome the stabilizing effect
of advection when $\chi < 3$. The minimum amplitude required for
a perturbation to rise buoyantly against the accretion flow is 
\citep{thompson00,scheck08,FT09b,dolence2013,couch2012}
\begin{equation}
\label{eq:delta_crit}
\left(\frac{\Delta \rho}{\rho}\right)_{\rm c} \simeq \frac{C_D\, v_2^2}{2l_v\, g_s} 
\end{equation}
where $v_2$ and $g_s$ are the postshock velocity
and gravitational acceleration at the shock, respectively, $C_D$ is the drag coefficient of the
perturbation ($\simeq 0.5$ for a sphere), 
and $l_v$ is the ratio of the volume to the cross-sectional area of the 
perturbation in the direction of gravity ($4/3$ times the radius, for a sphere). 

\begin{figure*}
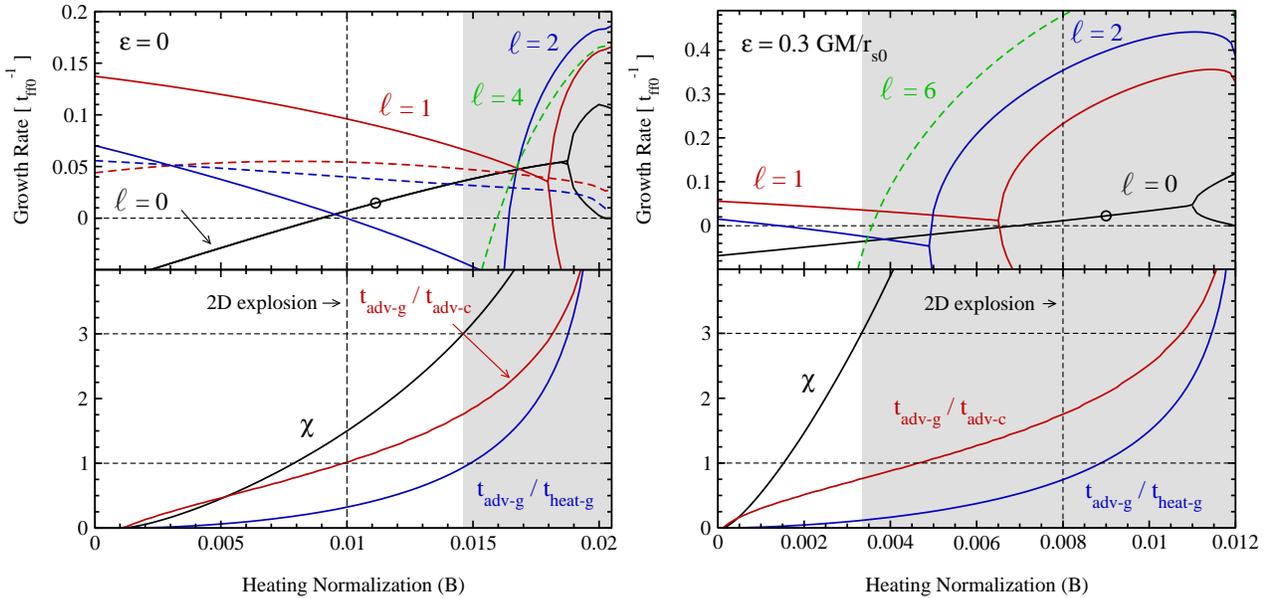

\begin{overpic}[width=0.97\columnwidth]{f1a.eps}
\put(18,63){\Large $\ell$}
\put(61,78.5){{\color{rplt}\Large $\ell$}}
\put(76,80){{\color{gplt}\Large $\ell$}}
\put(82,88.5){{\color{bplt}\Large $\ell$}}
\end{overpic}
\begin{overpic}[width=\columnwidth]{f1b.eps}
\put(20,64.5){{\color{rplt}\Large $\ell$}}
\put(40,78.7){{\color{gplt}\Large $\ell$}}
\put(70,87.5){{\color{bplt}\Large $\ell$}}
\put(78,63.5){\Large $\ell$}
\end{overpic}
\caption{Linear stability properties of initial models, as a function
of heating rate $B$ (eq.~[\ref{eq:neutrino_source_term}]-[\ref{eq:B_dimensional}]). 
Shown are SASI- and convection-dominated sequences (left and right, respectively). 
\emph{Top:} growth rates for $\ell=0,1,2$ modes in black, red
and blue, respectively, with
solid and dashed lines for fundamental mode and first overtone, respectively.
\emph{Bottom:} dimensionless parameters of the system evaluated in the
steady-state solution:
convection parameter $\chi$ (eq.~[\ref{eq:convection_parameter}], black), 
ratio of advection times in the gain and cooling regions (red), and ratio
of advection to heating times in the gain region (blue). 
The shaded area shows the region of parameter space where convection is
expected to overcome advection ($\chi > 3$, \citealt{foglizzo06}), open circles
signal the onset of $1D$ explosion (Appendix~\ref{s:L0}) and
vertical dashed lines the point of $2D$ explosion with 
the employed resolution (\S\ref{s:setup}). Green dashed lines
show the modes $\ell_{\rm crit}$ from equation (\ref{eq:lmax_def}).}
\label{f:growth_timescales}
\end{figure*}

We evolve two different sequences of models for which the heating rate
$B$ is varied from zero to a value that yields an explosion in 2D, and a third
set of models that explores the effect of large-amplitude perturbations on 
a SASI-dominated background state. All models are summarized in Table~\ref{t:models}.

The first sequence (e0) is such that all models are well within
the $\chi<3$ regime, corresponding to a SASI-dominated system. This 
background flow is obtained by setting the dissociation parameter $\varepsilon$ to zero.
The flow is initially perturbed everywhere with random cell-to-cell velocity fluctuations, with an amplitude $0.1\%$
of the local radial velocity.

The second sequence (e3) has the dissociation parameter set to $30\%$ of the gravitational
energy at the shock position without heating ($r=r_{\rm s0}$). 
The larger density jump yields smaller postshock velocities \citep{FT09a}, increasing
the value of $\chi$. Most of the models in this sequence lie in the $\chi>3$ regime,
and are therefore convection-dominated. This combination of parameters is the
same as used in one of the sequences of \citet{FT09b}. The same set of initial perturbations
as in the e0 sequence are used.

A third set of models (p0) has large-amplitude initial perturbations applied mostly to the exploding
model of the e0 sequence. We explore the effect of random cell-to-cell density perturbations
in the entire computational domain
with an amplitude of $10\%$ and $30\%$, overdense shells in the upstream flow that trigger specific SASI modes 
(see \citealt{FT09a} for details), and
density perturbations in the gain region which are radially-constant from $r_g$ to $r_s$, but with an 
angular dependence set by a Legendre polynomial. The latter
are aimed at exciting convection by large-scale perturbations. 
The amplitude is chosen to be $50\%$.

Note that we focus on exploding models that are marginally above the heating
rate for explosion, where non-radial instabilities are expected to have the maximum
effect. Further increase of the heating rate yields explosions that develop
earlier, eventually approaching the spherically-symmetric runaway 
condition (e.g., Appendix~\ref{s:L0}).

\subsubsection{Linear Stability Properties}
\label{s:linear_stability}

The linear stability properties of the two sets of background flow configurations (e0 and e3)
are shown in Figure~\ref{f:growth_timescales}. The growth rates of the fundamental $\ell=1$,
and $\ell=2$ modes as a function of heating rate are monotonically decreasing as long
as $\chi < 3$. Above $\chi > 3$, modes transition into a non-oscillatory (convective) state with two
branches, in line with the results of \citet{yamasaki07}. The mode $\ell_{\rm crit}$ that
bifurcates at the lowest heating rate ($\chi \simeq 3$) is approximately that for which $2\ell_{\rm crit}$ 
eddies of size $(r_s-r_g)$ fit into a transverse wavelength \citep{foglizzo06}
\begin{equation}
\label{eq:lmax_def}
\lambda_{\perp,\rm crit} \equiv \frac{\pi (r_s+r_g)}{\sqrt{\ell_{\rm crit}(\ell_{\rm crit}+1)}} \sim 2(r_s - r_g).
\end{equation}
Modes with larger or smaller $\ell$ bifurcate at higher heating rate.

Figure~\ref{f:growth_timescales} also shows two important timescale ratios in the
stationary solution as a function
of heating rates. The first one is the ratio of advection times in the gain and
cooling regions, $t_{\rm adv-g}$ and $t_{\rm adv-c}$, respectively. On the basis
of numerical simulations with a realistic EOS, \citet{F12} found that equality 
between these two timescales at $t=0$ corresponds approximately to the onset of
oscillatory instability. Figure~\ref{f:growth_timescales} shows that this relation
is valid for the $\varepsilon=0$ sequence, losing accuracy when nuclear dissociation
is included.

The \emph{instantaneous} value of the ratio of advection to heating timescales in the 
gain region has for long been used to quantify proximity to an 
explosion in numerical simulations \citep{janka98,thompson00,thompson05}.
\citet{F12} found that equality between these two timescales in the \emph{initial 
condition} -- or equivalently, at the time of shock stalling -- marks approximately
the subsequent onset of non-oscillatory $\ell=0$ instability in numerical simulations. 
However, Figure~\ref{f:growth_timescales} shows that the point where the linear
$\ell=0$ growth rate bifurcates to a non-oscillatory mode
lies at a much higher heating rate than the point where
$t_{\rm adv-g} = t_{\rm heat-g}$ in both sequences. Nevertheless, it is shown in 
Appendix~\ref{s:L0} that non-oscillatory instability still sets in at the heating
rate for which these timescales are equal in the initial condition, indicating 
that the expansion is a non-linear effect\footnote{For the e0 sequence, 
the e-folding time for the $\ell=0$ mode is approximately one half of the 
oscillation period at the heating rate for which $t_{\rm adv-g} = t_{\rm heat-g}$.}.

\subsection{Spherical Fourier-Bessel Spectral Decomposition}
\label{s:sfb_outline}

To analyze the properties of the flow accounting for its intrinsic spherical
geometry, we employ a spherical Fourier-Bessel expansion to perform various
spectral decompositions.
This set of functions forms an orthogonal basis of two- or three-dimensional 
space in spherical coordinates, allowing the expansion of an arbitrary 
scalar function $f(r,\theta,t)$ in a series of the form
\begin{equation}
\label{eq:sfb_2D}
f(r,\theta,t) = \sum_{n,\ell} f_{n\ell}(t) g_\ell(k_{n\ell}r) P_\ell(\cos\theta),
\end{equation}
where $g_\ell(k_{n\ell}r)$ are the radial basis functions, $k_{n\ell}$ is 
the radial wave number of order $n$, $P_\ell(\cos\theta)$ are the 
Legendre polynomials of index $\ell$, and $f_{n\ell}(t)$ are (time-dependent)
scalar coefficients. 
Expansions of this form have previously been used in the context of galaxy 
redshift surveys (e.g., \citealt{fisher1995}). Appendix~\ref{s:sfb_appendix} contains a
detailed description of the expansion method, including the straightforward 
extension to three-dimensional space. In what follows we provide a brief 
outline, focusing on the quantities needed to analyze the turbulent flow 
in our 2D models.

The domain considered is the volume enclosed between two concentric 
spheres of inner and outer radii $r_{\rm in}$ and $r_{\rm out}$, respectively.
These spherical surfaces can be any pairwise combination
of the neutrinosphere, gain radius, or shock radius, depending on the
particular region to be studied.
The radial basis functions $g_\ell(k_{n\ell}r)$ are linear combinations 
of spherical Bessel functions $j_\ell$ and $y_\ell$, with coefficients chosen 
to satisfy specific boundary conditions at both interfaces 
(Appendix~\ref{s:sfb_appendix}). 
 
Imposing these boundary conditions generates a set of
discrete radial wave numbers $k_{n\ell}$, in analogy with the modes of a membrane in
cylindrical coordinates. In addition to its quantum numbers $n$ and $\ell$, these
wave numbers depend on the chosen  ratio of inner and outer radii $r_{\rm in}/r_{\rm out}$.
Appendix~\ref{s:sfb_appendix} derives the wave numbers, relative coefficients, and
normalization of the radial basis functions for the cases of vanishing (Dirichlet)
and zero gradient (Neumann) boundary conditions. 
For low $n$, $\ell$, and 
$r_{\rm in}/r_{\rm out}\to 1$, these wave numbers approach
\begin{equation}
k_{n\ell}\to \frac{\pi}{(r_{\rm out}-r_{\rm in})}(n+1), \quad (n=0,1,2,...)
\end{equation}
increasing in value for stronger curvature.
The normalized basis functions
satisfy the orthogonality relation (equations~\ref{eq:normalization_condition_dirichlet} 
and \ref{eq:normalization_condition_neumann})
\begin{equation}
\int_{r_{\rm in}}^{r_{\rm out}}r^2\totd r\, g_{\ell}(k_{n\ell}r)g_{\ell}(k_{m\ell}r) = \delta_{nm}.
\end{equation}
The coefficients for the spherical Fourier-Bessel expansion in 
equation~(\ref{eq:sfb_2D}) are thus
\begin{equation}
f_{n\ell}(t) = \frac{2\ell+1}{2}\int f(r,\theta,t) g_\ell(k_{n\ell}r)\,P_\ell(\cos\theta)\,
r^2dr\,\sin\theta\totd\theta.
\end{equation}

From Parseval's identity, 
\begin{equation}
\label{eq:parseval_2d}
\int |f|^2\, \totd^2 x = \sum_{n,\ell}  \frac{2}{2\ell+1}\,|f_{n\ell}|^2,
\end{equation}
one can define a discrete power spectral density in 2D space
\begin{equation}
P_{n\ell} = \frac{2}{2\ell+1}\,|f_{n\ell}|^2.
\end{equation}
The coefficients $f_{n\ell}(t)$ can also be Fourier analyzed
in time, yielding an individual power spectrum for each $(n,\ell)$ 
mode. Using a Discrete Fourier Transform (DFT) in time, the normalization 
can be taken to be the time-average of the volume integral of the
variable in question (e.g., \citealt{NR}),
\begin{eqnarray}
\frac{1}{N_q}\sum_q \int |f|\, \totd^2 x & = & \frac{1}{N_q^2}\sum_{n\ell q} \frac{2}{2\ell+1}\,|\widehat{f}_{n\ell q}|^2\\
\label{eq:3d_spectrum}
&\equiv & \sum_{n\ell q} \mathcal{P}_{n\ell q},
\end{eqnarray}
where $N_q$ is the number of time samples, and $\widehat{f}_{n\ell q}$ is the DFT of 
$f_{n\ell}(t)$ at frequency $q$.

In practical applications, the series in equation~(\ref{eq:parseval_2d}) must be 
truncated at a finite value of the indices. In our analysis we set these maximum indices
to be at most half the number of grid points in the corresponding direction, in analogy
with the Nyquist limit in cartesian coordinates.

\subsection{General Relativistic, Radiation-Hydrodynamic Simulations}

We use the set of two-dimensional, 
general-relativistic, radiation-hyrodynamic simulations of \citet{mueller2012} to
test the validity of the general principles inferred from the parametric models.
The \citet{mueller2012} models follow the evolution of a star of mass  $8.1M_\odot$ and metallicity
$Z=10^{-4}$ (A. Heger 2013, private communication), and a $27M_\odot$ star of solar metallicity \citep{woosley02}.
The code employed is VERTEX-CoCoNuT \citep{mueller2010}, which treats multi-group 
neutrino transport using the `ray-by-ray-plus' approach \citep{rampp02,bruenn2006,buras06a}.

These two successfully exploding models follow very different paths on their way to
runaway expansion. The $8.1M_\odot$ progenitor (model u8.1) becomes dominated by convection 
shortly after the shock stalls, and remains so until runaway sets in.
In contrast, the $27M_\odot$ model (s27) develops a strong SASI throughout the
evolution.

\section{Results}
\label{s:results}

\subsection{Transition to Explosion}

We first concentrate on the differences in the transition to
explosion introduced by the initial dominance of the SASI or convection. 
To this end, we focus the discussion
on models that bracket the critical heating rate
for explosion (Table~\ref{t:models}). We then discuss the effect
of different initial perturbations on exploding models.

\begin{figure}
\includegraphics*[width=\columnwidth]{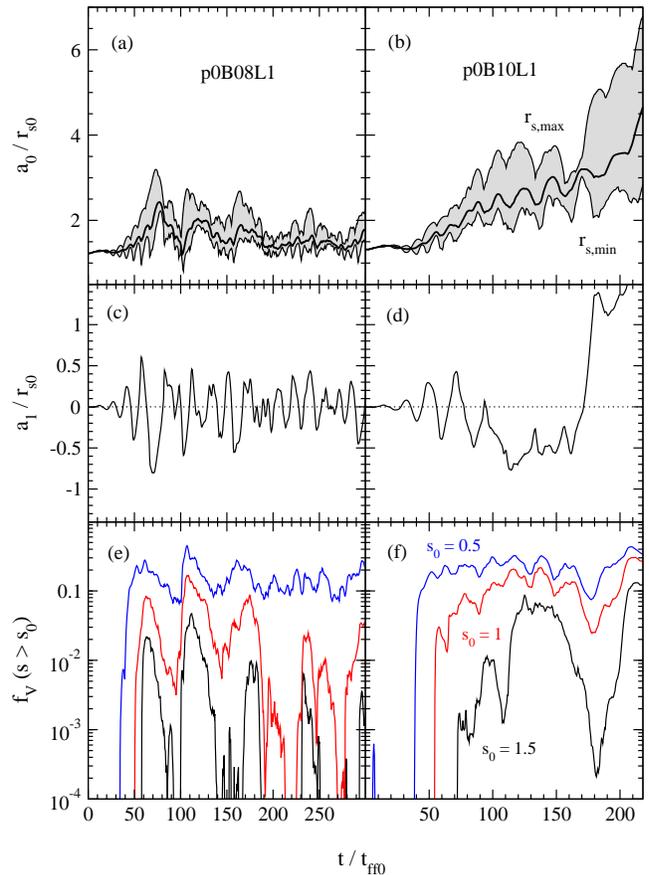}
\caption{Time-series diagnostics for SASI-dominated models below and above the
threshold for explosion (p0B08L1 and p0B10L1, respectively). The $\ell=1$ 
SASI mode is excited with an overdense shell. The top panels shows $\ell=0$ Legendre
coefficient (eq.~[\ref{eq:legendre_coeff}]) of the shock surface (thick line), 
as well as minimum and maximum shock radii (thin lines). Middle panels show
$\ell=1$ shock Legendre coefficient. Bottom panels show the fraction of the
post-shock volume with an entropy higher than a given value (eq.~[\ref{eq:volume_fraction}]).
Note that bubble destruction (sudden decreases
in $f_V$ for high entropy) precedes large amplitude sloshings of the 
shock (as indicated by $a_1$ changing sign). The unit of length
is the initial shock radius without heating $r_{\rm s0}$ and the unit of time is
the free-fall time at this position ($\sim 3$~ms for a central mass of
$1.3M_\sun$ and $r_{\rm s0}\sim 150$~km, \S\ref{s:setup}).}
\label{f:shock_entropy_sasi}
\end{figure}

\begin{figure}
\includegraphics*[width=\columnwidth]{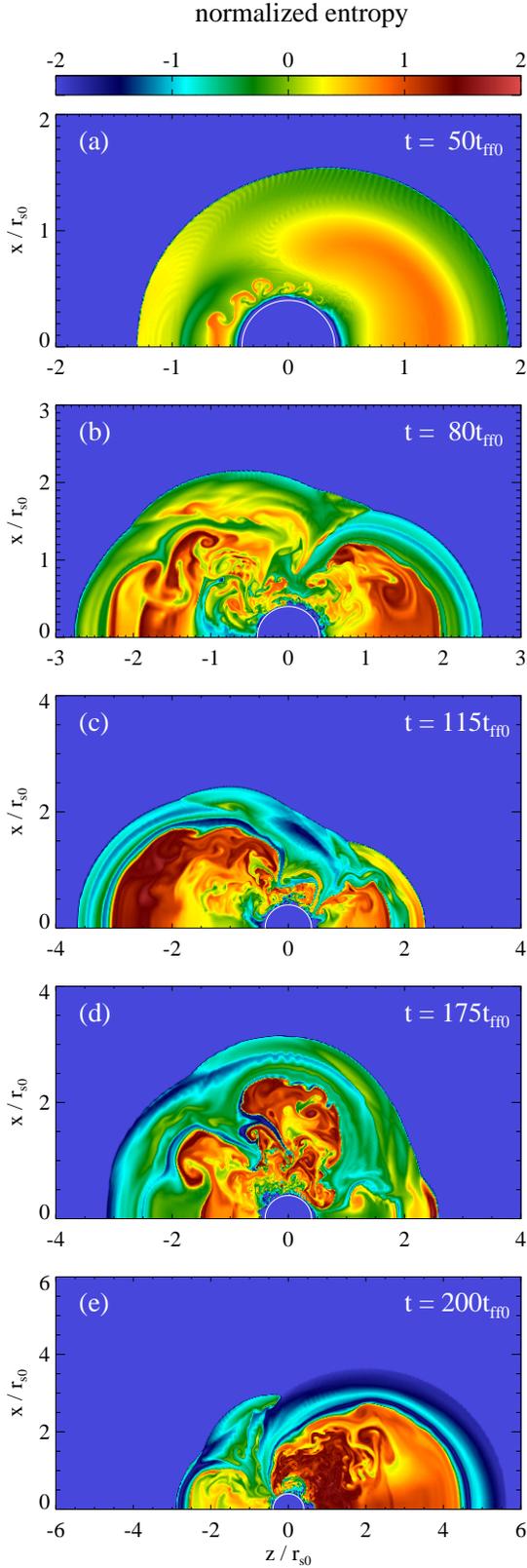}
\caption{Entropy at selected times in the evolution of model p0B10L1, which explodes SASI-dominated. 
Panels show the seeding of
perturbations in one hemisphere and bubble disruption on the other (a), loss of coherence of SASI (b), development
of first large bubble (c), partial disruption and displacement of bubble (d), and
final expansion (e). Compare with Figure~\ref{f:shock_entropy_sasi}.}
\label{f:shock_entropy_snapshots}
\end{figure}

\subsubsection{Interplay of SASI and Convection}

The characteristic behavior of models with an early dominance of the SASI 
is illustrated in Figures~\ref{f:shock_entropy_sasi} 
and \ref{f:shock_entropy_snapshots}. Initially, the $\ell=1$ shock 
Legendre coefficient displays sinusoidal oscillations of exponentially
growing amplitude. While in models without heating the SASI grows in
amplitude until oscillations saturate while keeping its characteristic
period (e.g., \citealt{FT09a}), in models with significant
heating this period increases when the amplitude becomes large, and
eventually the regularity of the oscillation is lost.

\begin{figure}
\includegraphics*[width=\columnwidth]{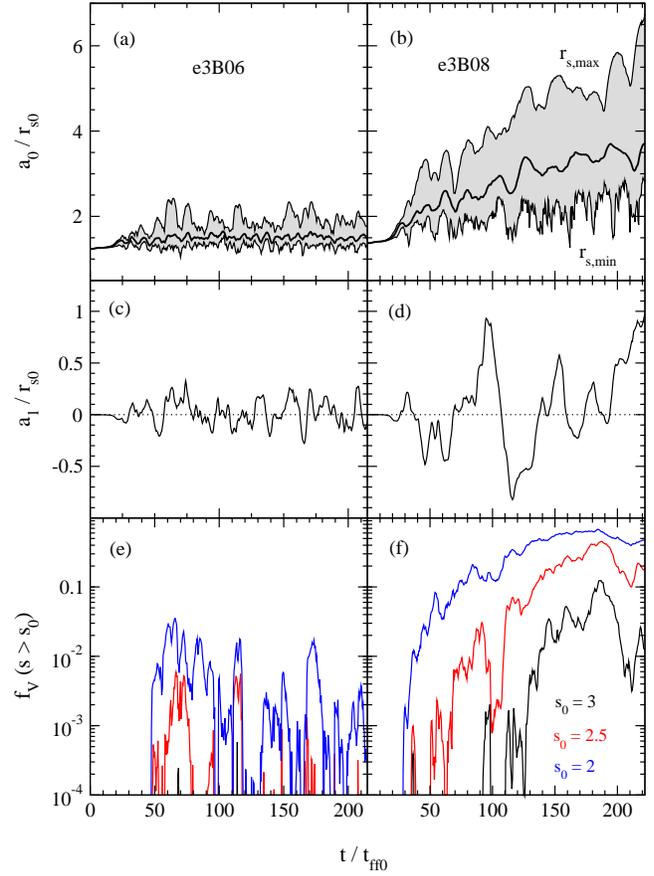}
\caption{Same as Figure~\ref{f:shock_entropy_sasi}, but for 
convection-dominated models that bracket the threshold
for explosion (e3B06 and e3B08). Even though $\ell=1$ shock oscillations
have no clear periodicity, the relation between destruction of high-entropy
bubbles and large amplitude shock sloshings is still present.
The unit of length
is the initial shock radius without heating $r_{\rm s0}$ and the unit of time is
the free-fall time at this position ($\sim 3$~ms for a central mass of
$1.3M_\sun$ and $r_{\rm s0}\sim 150$~km, \S\ref{s:setup}).}
\label{f:shock_entropy_conv}
\end{figure}

This breakdown of the SASI cycle is due to large-scale, long-lived fluid parcels
with enhanced entropy emerging in the post-shock region. These structures are seeded during shock 
expansions (Figure~\ref{f:shock_entropy_snapshots}a; see also \citealt{scheck08}). 
For small shock displacements, these elongated bubbles are shredded
by lateral flows inherent in the SASI, and are advected out of the gain region,
allowing the advective-acoustic cycle to proceed as in the case without heating. 
Above a certain amplitude, however, bubbles are able to resist shredding, 
and the SASI cycle is interrupted. Accretion proceeds then along narrow 
downflows that circumvent the bubbles (Figure~\ref{f:shock_entropy_snapshots}b).

To quantitatively analyze the interplay between sloshing of the post-shock region 
and large-scale bubbles, we compare in Figure~\ref{f:shock_entropy_sasi} 
the evolution of the $\ell=1$ shock Legendre coefficient $a_1$, where
\begin{equation}
\label{eq:legendre_coeff}
a_\ell(t) = \frac{2\ell+1}{2}\int_0^{\pi} r_s(\theta,t) P_\ell(\cos\theta)\sin\theta\,\totd \theta,
\end{equation}
with the fraction of the post-shock volume with entropy higher than
a fiducial value $s_0$:
\begin{equation}
\label{eq:volume_fraction}
f_V(s>s_0) = \frac{1}{V}\int_{s_0}^\infty \frac{\totd V}{\totd s}\totd s,
\end{equation}
where the entropy
\begin{equation}
s = \frac{1}{\gamma-1}\ln\left[\frac{p}{p_s}\left(\frac{\rho_2}{\rho}\right)^\gamma\right]
\end{equation}
is defined so that it vanishes below the shock in the initial model ($p_2$ and
$\rho_2$ are the initial post-shock pressure and density, respectively; e.g. 
\citealt{F07}), and the post-shock volume is defined as
\begin{equation}
V(t) = 2\pi\int_{0}^\pi\int_{r_*}^{r_s(\theta,t)}r^2\totd r\, \sin\theta\totd\theta.
\end{equation}
We use volume instead of mass to minimize the influence of low-entropy downflows.
The emergence of peaks in $f_V(t)$ for high values of the entropy is related
to the loss of periodicity and eventual halting of shock sloshings, while 
bubble destruction can allow regular periodicity to emerge again (c.f.
Figure~\ref{f:shock_entropy_sasi}c,e in the range $t\in [200,250]t_{\rm ff0}$).

Large-scale bubbles that have halted the SASI can nevertheless be broken 
when low-entropy downflows bend and
flow laterally. This process triggers bubble disruption, and
results in their shredding or displacement to the opposite
hemisphere (Figure~\ref{f:shock_entropy_snapshots}). Accretion is then
able to proceed through the whole hemisphere previously occupied by
the bubble, and the shock executes a sloshing (c.f. Figure~\ref{f:shock_entropy_sasi}d,f
in the range $t\in[150,200]t_{\rm ff0}$, also Figure~\ref{f:shock_entropy_snapshots}d). 
The shock retractions are related to a decrease in pressure 
support triggered by an increase in cooling. The buoyancy of 
high-entropy bubbles blocks the flow of gas to the cooling region, 
resulting in a lower amount of cooling per SASI cycle and a loss of 
periodicity in the shock oscillations.

The key difference between exploding and non-exploding models
appears to be whether the system can form entropy perturbations
of sufficient size and amplitude. Model p0B10L1 displays such an 
entropy enhancement at time $t\sim 115t_{\rm ff0}$. This enhancement is perturbed
and displaced around time $t\simeq 175t_{\rm ff0}$, triggering
a large sloshing of the shock that transitions into runaway expansion.
In contrast, model p0B08L1 fails to develop a long-lived structure
with entropy higher than $s>1.5$. The transition to explosion
for a large enough bubble results from the relative importance of buoyancy
and drag forces \citep{thompson00}.

The characteristic evolution of convection-dominated models is illustrated 
by Figure~\ref{f:shock_entropy_conv}. Entropy enhancements are 
initially generated by convection. Bubbles grow and merge into large-scale
structures, which cause non-linear shock displacements. In non-exploding
models, bubbles have a short lifetime, and hence the shock undergoes
sloshings of moderate amplitude over a range of temporal frequencies.
Note that the destruction of large bubbles can also lead to shock sloshings,
but the persistent generation of entropy fluctuations of smaller scale 
and amplitude prevent the emergence of SASI oscillations with a well-defined
periodicity.

For high enough heating rate, large bubbles are able to survive
for many eddy turnover times, leading to explosion in a manner 
similar to that of SASI-dominated models. The role of high-entropy
bubbles in convection-dominated models has been documented
previously \citep{dolence2013,couch2012}.

\begin{figure}
\includegraphics*[width=\columnwidth]{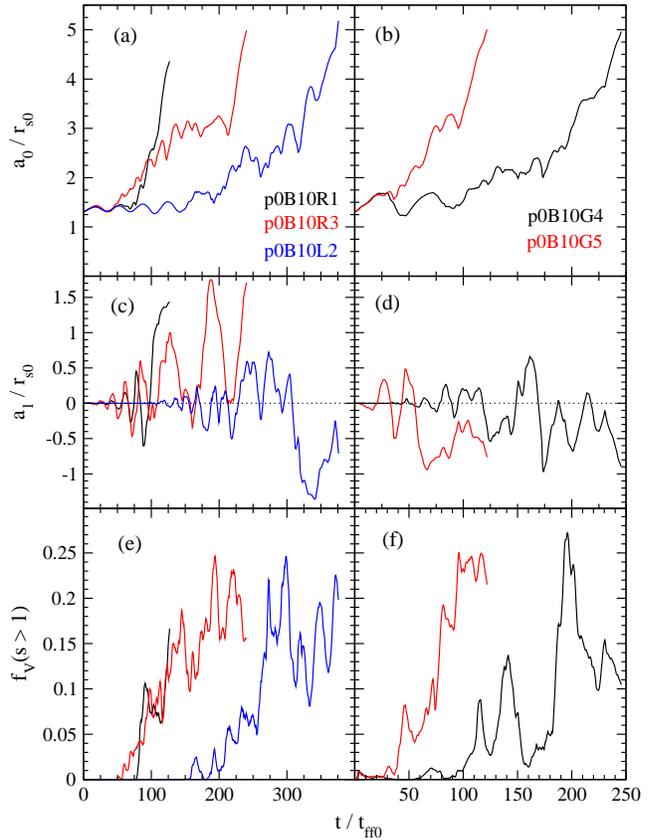}
\caption{Time-series diagnostics for exploding models with different
initial perturbations (Table~\ref{t:models}). Top panels show
average shock radius, middle panels show $\ell=1$ shock Legendre
coefficient, and bottom panel shows fraction of the postshock
volume with entropy higher than unity. Note the longer time to
explosion and late onset of $\ell=1$ oscillations in models
with even $\ell$ perturbations (p0B10L2 and p0B10G4).
The unit of length
is the initial shock radius without heating $r_{\rm s0}$ and the unit of time is
the free-fall time at this position ($\sim 3$~ms for a central mass of
$1.3M_\sun$ and $r_{\rm s0}\sim 150$~km, \S\ref{s:setup}).}
\label{f:shock_entropy_pert}
\end{figure}

\subsubsection{Effect of Initial Perturbations}

The effect of different initial perturbations on the exploding model
of the e0 sequence is illustrated in Figure~\ref{f:shock_entropy_pert}.
Models with large amplitude random cell-to-cell density perturbations (p0B10R1 and pB10R3)
follow the same path as the model where $\ell=1$ is directly
perturbed (p0B10L1, Fig.~\ref{f:shock_entropy_sasi}). The model
with an $\ell=2$ perturbation (p0B10L2) undergoes a weak convective
phase over a number of advection times, during which $\ell=0$
grows and $\ell=2$ saturates at a small amplitude.
After a delay of $\sim 100t_{\rm ff0}$, however, $\ell=1$
oscillations of the shock emerge, and the model joins the
usual SASI-dominated explosion path. The models with large amplitude
density perturbations in the gain region with a fixed $\ell=4$ and $5$ dependence
(p0B10G4 and p0B10G5) trigger less regular sloshings of the shock,
which however still result in the formation of large-scale bubbles.
As with the $\ell=2$ perturbation, an even-$\ell$ convective perturbation
takes longer to couple to an $\ell=1$ SASI mode.

The time to explosion appears to be a non-trivial function of the
perturbation form and amplitude. Model p0B10R3 has larger amplitude
perturbations, yet it hits the outer boundary $100$ dynamical times
later than model p0B10R1. Despite the very
large amplitude perturbation of model p0B10G4, it explodes later than
all models with an odd $\ell$ perturbation. This strong sensitivity
to initial conditions has been documented previously by \citet{scheck06}.

We emphasize however that we are focusing on models that are barely
above the threshold for explosion. Recently, \citet{couch2013c} have
pointed out the importance of pre-collapse perturbations in tilting
the balance towards explosion. Such an effect is likewise only going
to make a difference if a model is already close to exploding in the 
absence of perturbations. For instance, models e0B10 and p0B10L1 differ
in the type and amplitude of perturbations, leading to explosions that
differ by more than $100$ dynamical times in onset. In contrast, 
neither of models e0B08 or p0B08L1 explode, despite the fact that
they mirror the exact perturbations as the previous two exploding models
(the latter having a $10\%$ density perturbation in the form of a 
thin shell).

From our results it is not obvious that a large enough 
density perturbation suffices to turn a model for which the background 
state is SASI-dominated into a convectively dominated model. 
Note however that our models have $\chi \ll 3$. Previous
studies have witnessed more sensitivity to the type of 
initial perturbation when the $\chi$ parameter at shock stalling
is close to or even transiently exceeds criticality \citep{scheck08,hanke2013}.

\begin{figure}
\includegraphics*[width=\columnwidth]{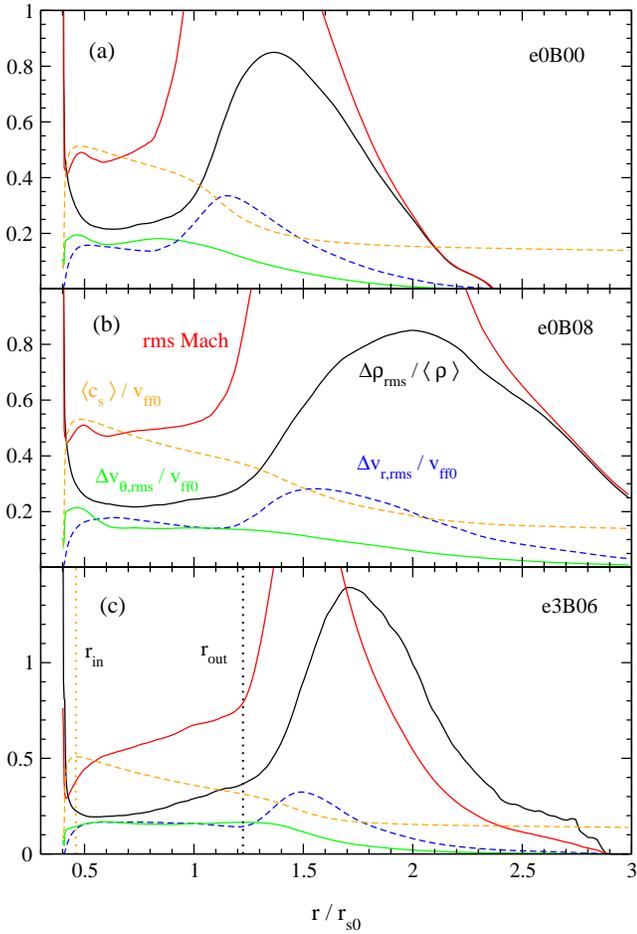}
\caption{Time- and angle-average profiles of selected quantities for non-exploding
models. Top, middle, and bottom panels show models with no heating (e0B00), SASI-dominated
close to explosion (e0B08), and convection-dominated close to explosion (e3B06), respectively.
Curves correspond to r.m.s. density fluctuation normalized to its mean value at each radius
(eqns.~[\ref{eq:time_average}]-[\ref{eq:rms_fluctuation}], solid black), 
r.m.s. Mach number (solid red), r.m.s. radial velocity (dashed blue),
r.m.s. meridional velocity (solid green), and average sound speed (dashed orange).
The vertical dotted lines in panel (c) bracket the radial range where the post-shock
flow is subsonic and free from strong stratification effects, with $r_{\rm in}$ and
$r_{\rm out}$ corresponding to the peak of the average sound speed, and the
average of the minimum shock radius minus its r.m.s. fluctuation 
(c.f. Fig. 8 of \citealt{FT09a}), respectively.}
\label{f:profiles_timeave}
\end{figure}

The `purity' of an excited $\ell=1$ SASI mode also
depends on whether the background flow allows for
unstable harmonics. Figure~\ref{f:growth_timescales} shows
that the first $\ell=1$ overtone is unstable for all
the heating rates in the e0 sequence. This may lead to shock oscillations
that are not a clean sinusoid, but which should not be mistaken
as an imprint of convection.

\subsection{Properties of the Quasi-Steady State}
\label{s:quasi-steady}

We now address the properties of the turbulent flow
in the gain region in cases where an explosion is not
obtained, focusing on the differences between models where
either SASI or convection dominate.
We first discuss general properties of the time-averaged
flow, and then analyze models using a spherical Fourier-Bessel
decomposition in space and a discrete Fourier transform in time.

\subsubsection{Time-Averaged Flow and Convective Stability}

\begin{figure}
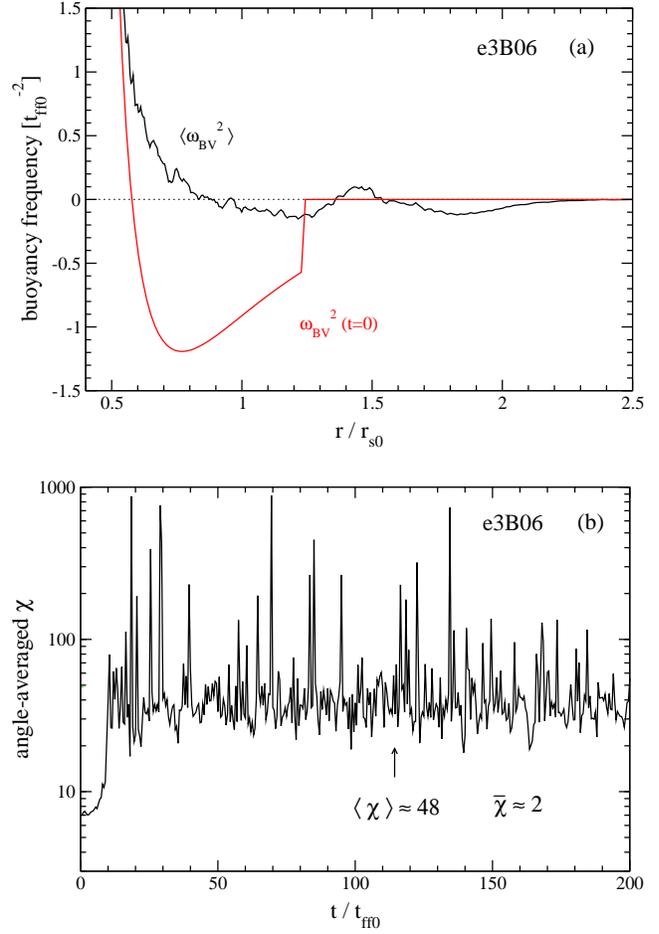

\includegraphics*[width=\columnwidth]{f7a.eps}
\includegraphics*[width=\columnwidth]{f7b.eps}
\caption{\emph{Top:} Squared buoyancy frequency (eq.~[\ref{eq:brunt}]) as a function of radius for
a convection-dominated model close to explosion (e3B06). Curves show initial
(red) and time-angle-averaged values (black). \emph{Bottom:} angle-averaged
convection parameter as a function of time for the same model (e3B06).
The time average value $\langle \chi\rangle$ is much larger than what is
obtained when computing this parameter with quantities from the time-averaged flow,
$\bar\chi$ (eq.~[\ref{eq:chi_mean_flow}]), because $\chi$ is a non-linear function.}
\label{f:brunt_comparison}
\end{figure}

The spatial structure of the quasi-steady-state becomes clear
when the flow is averaged in angle and time (e.g., \citealt{FT09a}).
Figure~\ref{f:profiles_timeave} shows such a representation
for a model without heating (e0B00), as well as SASI- and convection-dominated
models close to an explosion (e0B08 and e3B06, respectively).
The time- and angle-average of a generic scalar quantity is
denoted by
\begin{equation}
\label{eq:time_average}
\langle A(r) \rangle = \frac{1}{2(t_{\rm f}-t_{\rm i})}\int_{t_{\rm i}}^{t_{\rm f}}\totd t\,
                    \int_0^\pi A(r,\theta,t)\,\sin\theta\totd\theta,
\end{equation}
where $[t_i,t_f]$ is the time interval considered for the average, 
and the corresponding root-mean-square (r.m.s.) fluctuation is defined as
\begin{equation}
\label{eq:rms_fluctuation}
\Delta A_{\rm rms} = \left[ \langle A^2\rangle-\langle A\rangle^2\right]^{1/2}.
\end{equation}

All three models share a basic general
structure. From the inside out, this structure is composed of a narrow 
cooling layer adjacent to $r_*$, a region of 
sub-sonic turbulence encompassing part of the cooling layer and 
part of the (time-averaged) gain region, an extended zone 
of shock oscillation,
and the unperturbed upstream flow.

The most notorious difference among these models lies in the properties
of the shock oscillation 
zone and in the flow around the cooling
layer. Models where the SASI dominates have a wider shock oscillation
zone than the model where convection is dominant. This can be seen
by comparing the minimum and maximum shock radii of the non-exploding models
in Figures~\ref{f:shock_entropy_sasi} and \ref{f:shock_entropy_conv}. 
Also, in models where the SASI is prominent there is a bump in the r.m.s
lateral velocity in the cooling layer, indicating strong shear.
This bump is absent in the convection-dominated model.

In contrast, the subsonically turbulent 
region has very similar properties
in the three different models shown in Figure~\ref{f:profiles_timeave}, with
only slight changes in the radial slopes.
Characteristic values are $\Delta \rho_{\rm rms}/\langle \rho\rangle \sim 0.25$,
r.m.s. Mach number $\sim 0.5$, and 
$\Delta v_{\rm r,rms}\simeq \Delta v_{\theta,{\rm rms}}\sim 0.15v_{\rm ff0}$.
This similarity in time-averaged properties suggests that flows are not
very different from each other.

By analogy with convective systems in steady-state (e.g., nuclear burning
stars), one can investigate whether the time-averaged system adjusts
itself to a state of marginal convective stability. In hydrostatic
systems, convection acts to erase destabilizing gradients, whereas the presence
of advection in core-collapse supernova flows generates a non-zero
entropy gradient in steady-state \citep{murphy11}. One can nevertheless
ask whether the relevant critical parameter for convection is restored to stability
in the non-linear regime.

\begin{figure*}
\begin{overpic}[width=\textwidth]{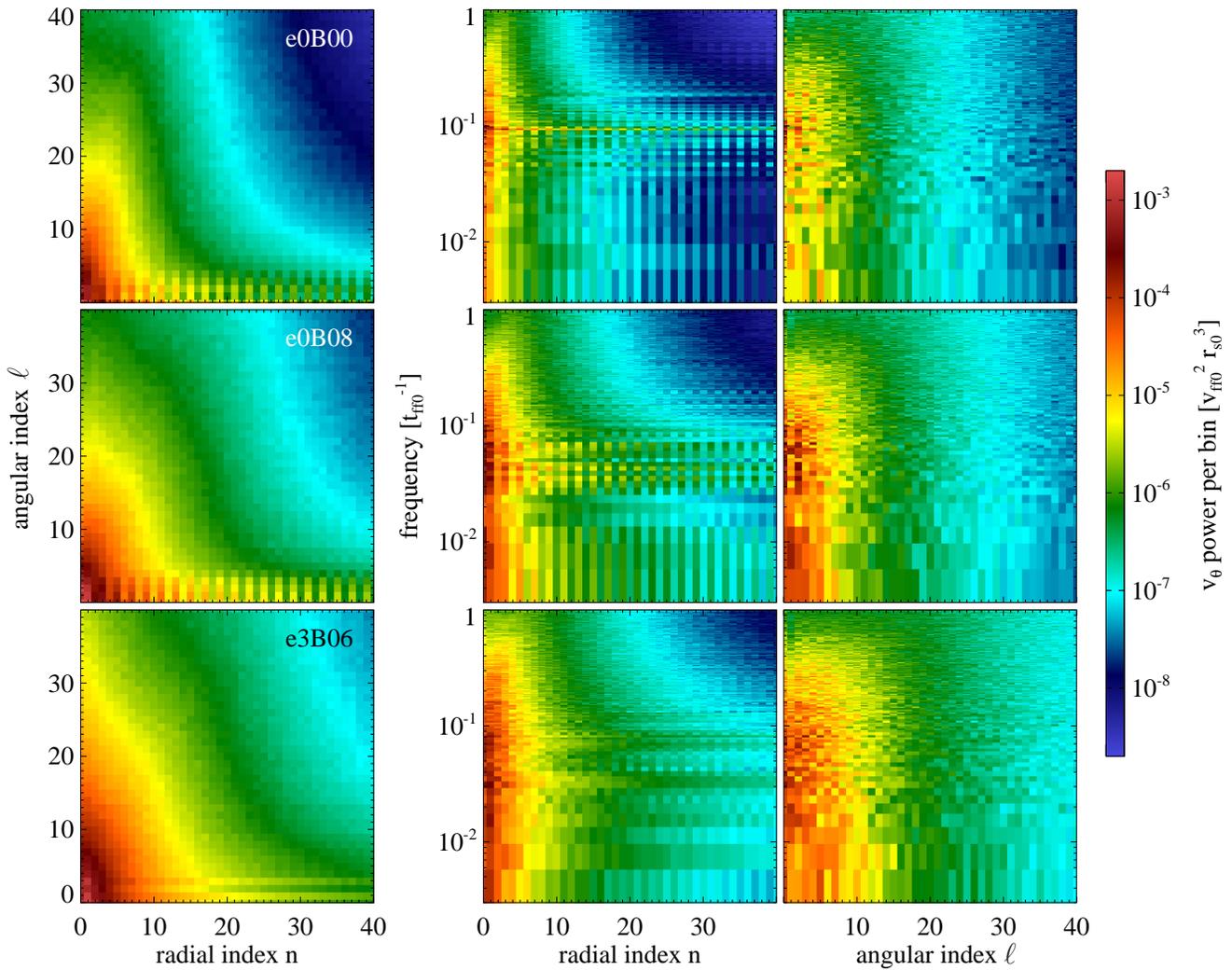}
\put(81.5,1){\Large $\ell$}
\put(1.3,48){\rotatebox{90}{\Large $\ell$}}
\end{overpic}
\caption{Two-dimensional spherical Fourier-Bessel power spectra, obtained by contracting 
the three-dimensional space-time power $\mathcal{P}_{n\ell q}$ (eq.~[\ref{eq:3d_spectrum}]) 
along one dimension, for
a model with no neutrino heating (e0B00, top row), a SASI-dominated model close to explosion
(e0B08, middle row), and a convection-dominated model close to explosion (e3B06, bottom row). 
Shown are frequency-summed spatial spectra (left column), $\ell$-summed time-$n$ spectra (middle
column), and $n$-summed time-$\ell$ spectra (right column). Models with a strong SASI display
an even-odd pattern in the radial direction at low $\ell$, and enhanced power near the advection 
frequency $\sim 0.1t_{\rm ff0}^{-1}$.}
\label{f:sfb_2d_mosaic}
\end{figure*}

\begin{figure*}
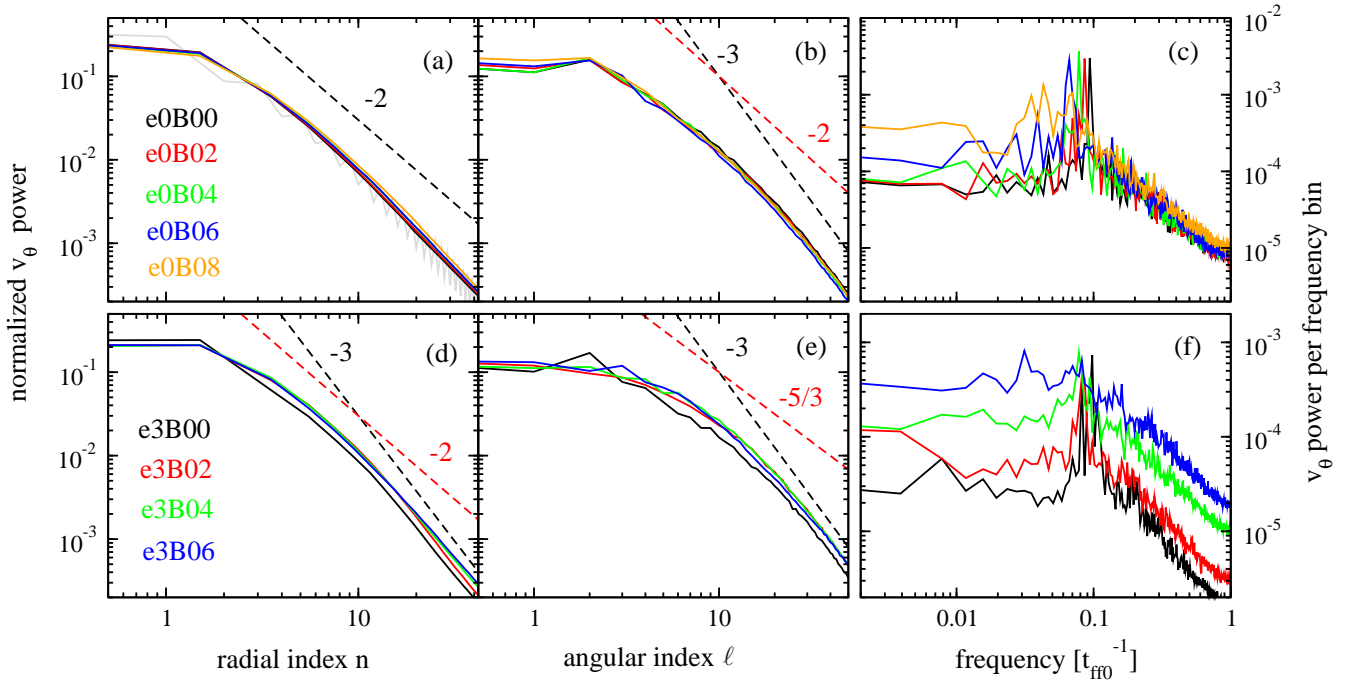

\begin{overpic}[width=\textwidth]{f9.eps}
\put(54,1.4){{\large $\ell$}}
\end{overpic}
\caption{One dimensional spectra for non-exploding models
(Table~\ref{t:models}), obtained by contracting $\mathcal{P}_{n\ell q}$
(eq.~[\ref{eq:3d_spectrum}]) along two dimensions. The top row shows
models in the SASI-dominated sequence (e0) and the bottom row contains
convection-dominated models (e3), as labeled. Left, middle, and right columns 
display radial, angular, and temporal spectra, respectively. The radial spectrum is smoothed
by averaging contiguous values; for reference, the unsmoothed spectrum for model
e0B00 is shown by the gray curve in panel (a). Dashed lines show reference
spectral slopes, labeled by the corresponding exponent.}
\label{f:spectra1d_mosaic}
\end{figure*}

Figure~\ref{f:brunt_comparison}a shows the initial and time-averaged squared buoyancy frequency 
(eq.~[\ref{eq:brunt}]) for model e3B06, which is convection dominated. This model
has an initial value of $\chi \simeq 7$ (Table 1). The time-averaged flow is such
that the degree of convective instability (negative $\omega_{\rm BV}^2$) is significantly
weaker than that in the initial state. The implications for convective
stability become clear when the $\chi$ parameter (eq.~[\ref{eq:convection_parameter}]) 
is computed for the time-average
flow. One way of doing this is simply averaging $\chi$ in time and angle, $\langle \chi\rangle$. 
However, because this is a non-linear function of the flow variables, the resulting value
will not only capture the properties of the mean flow, but it will also include the
contribution of turbulent correlations in the pressure, density, and velocity.
One can nevertheless still define a convection parameter based on the properties of the 
mean flow 
\begin{equation}
\label{eq:chi_mean_flow}
\bar{\chi} = \int \frac{{\rm Im}\left(\langle \omega_{\rm BV}^2\rangle^{1/2}\right)}{\langle v_{\rm r}\rangle}\,\totd r,
\end{equation}
where the integral extends over regions where $\omega_{\rm BV}^2 <0$.

The difference between these two ways of computing $\chi$ is illustrated 
in Figure~\ref{f:brunt_comparison}b. Shown is the instantaneous angle-averaged
value of $\chi$, together with its time average $\langle \chi\rangle$ as well
as the convection parameter computed using the mean flow, $\bar\chi$ (eq.~[\ref{eq:chi_mean_flow}]).
The instantaneous angle-averaged value of $\chi$ achieves very large values as soon
as the shock displacement becomes non-linear, similar to the results of
\citet{burrows2012}, with a time-averaged value $\langle \chi \rangle \sim 50$. The
convection parameter from the mean flow is much smaller, however, yielding
$\bar\chi \simeq 2$. This small number arises from the small magnitude
of the time-average of the squared buoyancy frequency shown in Figure~\ref{f:brunt_comparison}a.

Values of $\bar\chi$ for all non-exploding models are shown in Table~\ref{t:models}.
All convection-dominated models
satisfy $\bar\chi <3$, which indicates that in quasi-steady-state they adjust to
a state of convective sub-criticality (the equivalent of `flat' entropy gradients in
hydrostatic systems).
The SASI-dominated models maintain $\chi_0 \lesssim \bar{\chi} < 3$, where
$\chi_0$ is the value of $\chi$ in the initial condition. This \emph{increase}
in the time-averaged value of the convection parameter can arise
from the increase in the size of the gain region caused by SASI activity, and
from the presence of localized entropy gradients induced by the SASI, which
trigger secondary convection (e.g., Figure~\ref{f:shock_entropy_snapshots}a).

It is worth emphasizing that the driving agent matters in characterizing
convective motions: secondary convection is qualitatively different
from neurino-driven convection in that in the former there are both preferred spatial
and temporal scales (entropy perturbations induced by the SASI, and advection
time, respectively).

Another aspect of the explosion mechanism that can be probed with the time-averaged flow is the 
dependence of the turbulent kinetic energy in the gain region on neutrino heating.
\citet{hanke2012} found that a good indicator of the proximity of an explosion is the
growth of the turbulent kinetic energy on the largest spatial scales. Since the mass in 
the gain region also increases due to the larger average shock radius, it is worth clarifying
the origin of the increase in the kinetic energy. Table~\ref{t:models} shows the ratio
of the total time-averaged turbulent kinetic energy in the gain region to the time-averaged
mass in the gain region for non-exploding models. SASI-dominated models are such that
this ratio is nearly constant, decreasing slightly when an explosion is closer. Thus 
larger kinetic energy is due solely to the increase in the mass of the gain region. In contrast,
convection-dominated models grow both the specific kinetic energy and the mass in the
gain region as an explosion is closer.

\subsubsection{Properties of Turbulence in the Subsonic Region}

We now use the spherical Fourier-Bessel expansion to analyze the properties
of the turbulence in the subsonic region of the time-averaged flow. Operationally,
we define the radial limits of this region ($r_{\rm in}$ and $r_{\rm out}$, \S\ref{s:sfb_outline}) 
to be the peak
of the time-averaged sound speed, $\langle c_s\rangle$, and the time-average
of the minimum shock radius minus its r.m.s. fluctuation, 
$r_{\rm out} = \langle r_{\rm s,min}\rangle-\Delta r_{\rm s,min,rms}$, respectively.
This definition differs from that of \citet{murphy11} in that we restrict
ourselves to radii below the minimum shock position to avoid supersonic flow.

To connect with previous studies, we use the meridional velocity $v_\theta$ as
a proxy for the turbulent flow. We do not multiply this velocity by $\sqrt{\rho}$, however,
because the density stratification over the extended radial range considered 
would affect the spectral slopes \citep{endeve2012}. Thus, the sum of the total power (eq.~[\ref{eq:3d_spectrum}])
does not approach the total kinetic energy in the subsonic region, but instead it is
a measure of the kinetic energy per unit mass.

Figure~\ref{f:sfb_2d_mosaic} shows 2D projections of the 3D space-time spectrum $\mathcal{P}_{n\ell q}$
(eq.~[\ref{eq:3d_spectrum}]), for pure SASI, SASI-dominated, and convection-dominated models 
(c.f. Figure~\ref{f:profiles_timeave}).
Power is maximal at low angular and radial scales, as expected
from the inverse turbulent cascade in 2D (e.g., \citealt{davidson}). Models where the SASI
is prominent display two characteristic features: (1) an even-odd pattern in the 
radial spectrum for $\ell = 0-5$, indicating the presence of discrete modes, 
and (2) enhanced power around the frequency corresponding
to the advection time of the mean flow, $\bar{f}_{\rm adv}\sim 0.1t_{\rm ff0}^{-1}$.
The dominance of convection manifests as a broadening of the smoother component
of the spatial spectrum to larger $n$ and $\ell$, a near disappearance of the
even-odd pattern, and the emergence of power at temporal frequencies below and above $\bar{f}_{\rm adv}$.
This behavior of SASI- and convection-dominated models in the frequency-domain 
is consistent with the results of \citet{mueller2012} and \citet{burrows2012}.

Figure~\ref{f:spectra1d_mosaic} shows the results of contracting the
$\mathcal{P}_{n\ell q}$ array along two dimensions, yielding
one dimensional spectra, for models that do not explode.
In SASI-dominated models, the normalized power as a function of $n$ 
shows a characteristic sawtooth shape, which is smoothed to clarify the slope (an example of
an non-smoothed spectrum is shown by the gray curve in Figure~\ref{f:spectra1d_mosaic}a). 

Increasing the heating rate leads to minor changes in the (normalized)
radial spectrum in SASI-dominated models. The onset of convection, on the
other hand, leads to a shift of power from $n \leq 2$ to $n\geq 3$.
The spectral slope at large $n$ is approximately $n^{-2}$. This slope could be attributed
to Rayleigh-Taylor turbulence, for which the velocity fluctuations
satisfy $\delta v \propto \lambda^{1/2}$, with
$\lambda$ the wavelength of the perturbation
(e.g., \citealt{niemeyer1997,ciaraldi2009}). Note however that the 
wave numbers of the radial basis functions of different $\ell$
are not harmonic with each other (Fig.~\ref{f:zeroes_eigenfunctions_dirichlet}), 
hence one cannot straightforwardly map radial wavelength into index $n$.
Nevertheless, the spacing between wave numbers becomes nearly constant
at large $n$, with only a linear shift with $\ell$, motivating the use
of $n$ as a differential measure of the turbulent cascade.

The angular spectrum in SASI-dominated models shows a peak at $\ell=2$, and 
a slope at large $\ell$ indicative of a direct vorticity cascade \citep{kraichnan1967}.
Similar to the radial spectrum, the onset of convection results in the shift of 
power from $\ell \leq 2$ towards $\ell = 5-10$. The resulting spectral shape
has a form similar to that found by \citet{hanke2012}, \citet{couch2012}, and \citet{dolence2013}, 
who radially averaged the kinetic energy over a thin slice. This shape consists
of a shallow curved shape at low $\ell$, transitioning to  $\sim\ell^{-3}$
slope at large $\ell$.

The temporal spectrum of the sequence of convection-dominated models is consistent
with the results of \citet{burrows2012}. At very low heating rates, a prominent
peak exists at the advection frequency $\bar{f}_{\rm adv}$, indicating the
presence of the SASI. As the heating rate is increased, power increases
at frequencies below and above the advection peak. At heating rates close to an explosion,
this low-frequency power is comparable or higher than that at $\bar{f}_{\rm adv}$.

In contrast, the SASI-dominated sequence has a dominant peak at the advection
frequency for all models. This peak moves to lower frequencies as heating is
increased, because the advection time increases given the larger average shock radius (Table~\ref{t:models}).
Also, the peak becomes broader as a likely result of secondary convection being
triggered by the SASI.
Power at the lowest frequencies still increases with heating rate, but it remains below that
in the advection peak by at least a factor of two (in contrast, neutrino-driven
convection yields a nearly flat spectrum).
Note also that the power at frequencies higher than the advection peak 
in model e0B08 (SASI-dominated model with the highest heating) is within a 
factor of two of the convection-dominated model with the highest heating (e3B06).

From Figure~\ref{f:profiles_timeave} one can infer the turnover time of large eddies
to be $t_{\rm eddy} \sim 2\pi r/\Delta v_{\theta,{\rm rms}}\sim 30t_{\rm ff0}$,
yielding a frequency $f_{\rm eddy}\sim 0.03t_{\rm ff0}^{-1}$. Thus, the increase in power at frequencies below the
advection time appears to be associated with the evolution of large bubbles
in the gain region.

\subsection{Application to Full-Scale Core-Collapse Models}

\begin{figure}
\includegraphics*[width=\columnwidth]{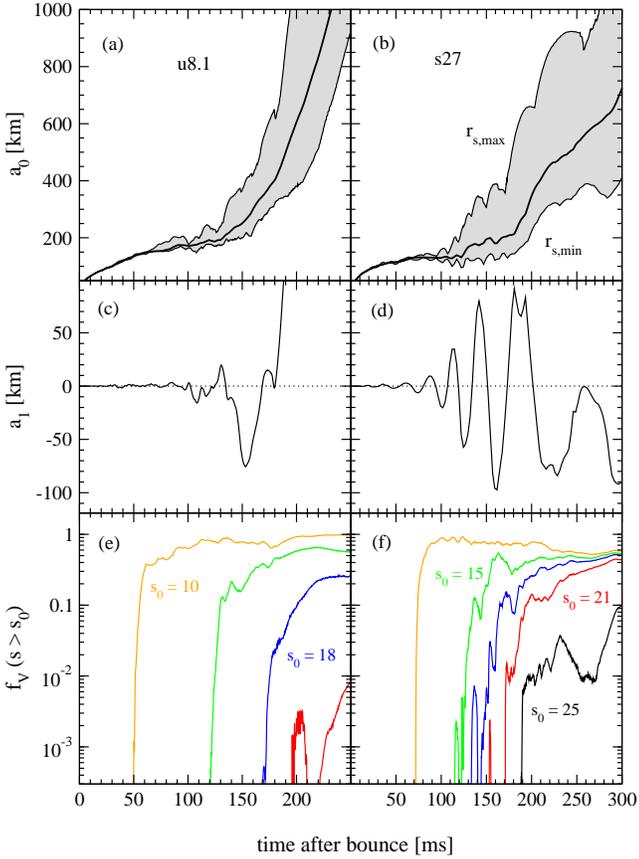}
\caption{Same as Figure~\ref{f:shock_entropy_sasi}, but for models u8.1
and s27 of \citet{mueller2012}. The fiducial entropies $s_0$ are in units
of $k_{\rm B}$ per baryon.}
\label{f:shock_entropy_mpa}
\end{figure}

\begin{figure}
\includegraphics*[width=\columnwidth]{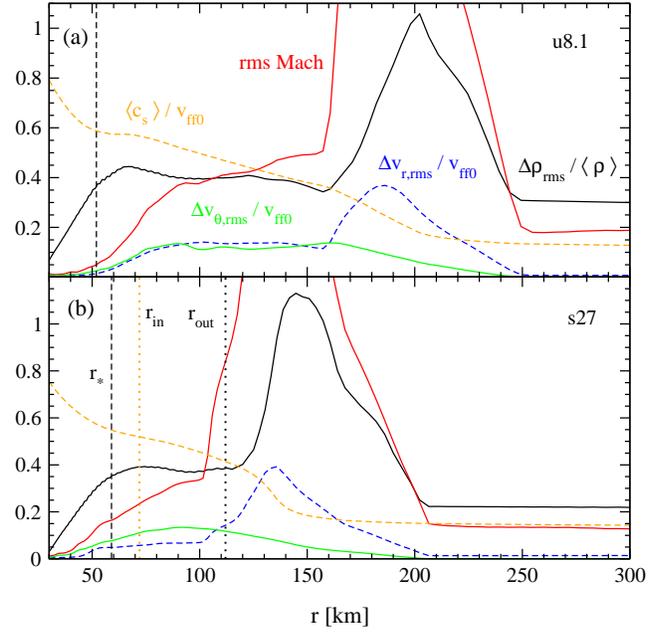}
\caption{Time- and angle-averaged profiles of selected quantities for
models u8.1 and s27 of \citet{mueller2012} (compare with Fig.~\ref{f:profiles_timeave}). 
The free-fall velocity normalization is computed in Newtonian gravity, for gravitational
masses $\{1.2,1.35\}M_\sun$ and initial shock radii $\{150,130\}$~km for
models \{u8.1,s27\}, respectively. The vertical dashed lines correspond to the
time- and angle-averaged radius for which $\rho = 10^{11}$~g~cm$^{-3}$, which we
associate with $r_*$. The vertical dotted lines bracket the radial extent of the
region used for spectral analysis (see text for details).}
\label{f:profiles_timeave_mpa}
\end{figure}

Here we analyze the models of \citet{mueller2012} with the same methods
used in our parametric models, identifying similarities and differences.

Figure~\ref{f:shock_entropy_mpa} shows the evolution of the $\ell=0$ and $\ell=1$
coefficients for models u8.1 and s27, together with the fraction of the volume
with entropy higher than fiducial values $s_0 = \{10,15,18,21,25\}~k_{\rm B}$ per baryon.
The $f_V$ diagnostic behaves similarly to exploding parametric models p0B10L1 and e3B08 
(Figs.~\ref{f:shock_entropy_sasi} and \ref{f:shock_entropy_conv}).
After a large enough fraction of the postshock volume is occupied by high entropy material,
the regular periodicity of shock oscillations in model s27 is modified ($t\simeq 150$~ms). Shock
sloshings in this late stage are preceded by partial disruption of bubbles. One notable
difference with model p0B10L1 is the emergence of secondary shocks in model s27, which
prevent complete disruption of high-entropy bubbles. Runaway expansion in model s27
is preceded by accretion of the Si/O composition interface. Another important difference 
between both \citet{mueller2012} models and the exploding parametric models is the level
of $\ell=0$ oscillations, which is much larger in the exploding gamma-law simulations\footnote{The difference
in shock expansion rate once runaway starts is due to the absence of alpha particle recombination in the
parametric models \citep{FT09b}; this is independent of the level of $\ell=0$ oscillations.}.

Models u8.1 and s27 both undergo a quasi-stationary phase that precedes
runaway expansion. We have analyzed the properties of the time-averaged
flow over the interval $[80,130]$~ms and $[70,120]$~ms in models u8.1
and s27, respectively. During these intervals, both the average shock
radius and the average neutrinospheric radius $r_*$ (defined as the isodensity
surface $\rho=10^{11}$~g~cm$^{-3}$) change by less than $20\%$. Figure~\ref{f:profiles_timeave_mpa}
shows the resulting profiles of time-averaged quantities, in analogy 
with Figure~\ref{f:profiles_timeave}. Above the neutrinosphere, all
quantities behave in the same qualitative way as the parametric models.
At densities $\rho=10^{11}$~g~cm$^{-3}$ and higher, clear differences
are introduced by the existence of a protoneutron star, however. In particular,
the density and velocity fluctuations decrease significantly inside
$r_*$, whereas Figure~\ref{f:profiles_timeave} shows a strong increase
in the density perturbation near $r_*$ for parametric models due to the accumulation of mass
given the reflecting boundary condition, and a bump in the lateral
velocity due to shear in SASI-dominated cases. 
Nonetheless, the very similar behavior of the system outside $r_*$
shows that a reflecting boundary condition is not a bad approximation.

\begin{figure*}
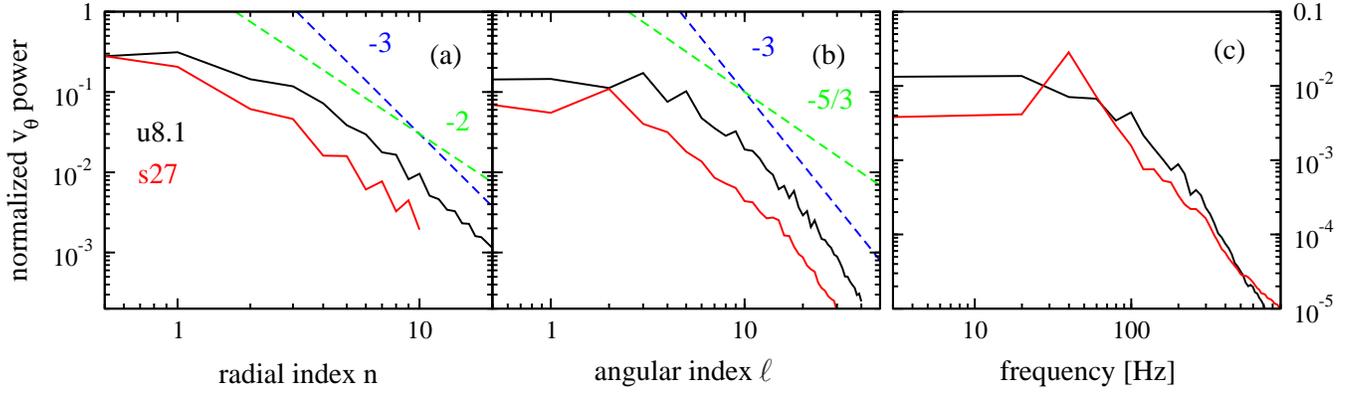

\begin{overpic}[width=\textwidth]{f12.eps}
\put(56.7,0.7){\Large $\ell$}
\end{overpic}
\caption{Same as Figure~\ref{f:spectra1d_mosaic}, but for models u8.1 and s27
 	 of \citet{mueller2012}. The temporal spectrum is normalized so that
	 the frequency integral is unity.}
\label{f:spectra1d_mosaic_mpa}
\end{figure*}

We have also computed the convection parameter using the time-averaged
flow (eq.~[\ref{eq:chi_mean_flow}]). The buoyancy frequency is computed
following \citet{mueller2013} but ignoring relativistic
corrections\footnote{The leading order corrections to the buoyancy
frequency and sound speed scale like $(c_s/c)^2$ \citep{mueller2013}, which is only
a few percent in the gain region.}. Model u8.1 has $\bar\chi \simeq 0.7$, consistent
with the hypothesis that convection-dominated flow adjusts itself to 
sub-criticality. This parameter is even smaller ($\bar\chi \simeq 0.4$) in model s27.

Figure~\ref{f:spectra1d_mosaic_mpa} shows one-dimensional spectra of the subsonic
region in models u8.1 and s27. The limits of the region are defined to be
the saddle point in the time- and angle-averaged sound speed on the inside,
and the time-average of the minimum shock radius minus its rms fluctuation on the
outside (both radii are shown in Figure~\ref{f:profiles_timeave_mpa} for s27).
The radial spectrum of the \citet{mueller2012} models has relatively less
power at long wavelengths than the parametric models. At short wavelengths,
however, the spectral slope is similar, and the sawtooth pattern of SASI-dominated
models is also present in model s27 (the spectrum extends to $n=10$ due to
the compactness of the subsonic region relative to the grid spacing). The angular
spectrum of model s27 shows the same peak at $\ell=2$ as the SASI-dominated
parametric models. Model u8.1 shows a similar curved shape as convection-dominated
models, though with some excess at $\ell = 3$ and $\ell=5$. Finally, the temporal
spectrum shows clearly the distinction between convection-dominance manifesting
more power at low frequencies (u8.1), and SASI-dominance generating a clear
peak at the advection time (s27). This temporal frequency behavior was already
noted by \citet{mueller2012}.

In summary, the general results from parametric models regarding SASI- and convection-dominated
flow persist in sophisticated, full-scale core-collapse simulations. The use of
nuclear dissociation in parametric models as the control parameter for switching
between SASI- and convection-dominance does not prevent 
qualitative agreement with full-scale models, even though the latter always 
have dissociation present. This is because the behavior of the flow depends
chiefly on the \emph{relative timescales} of the system 
(c.f. \S\ref{s:linear_stability}).

\section{Summary and Discussion}
\label{s:summary}

We have analyzed the transition to explosion in SASI- and convection-dominated
core-collapse supernova explosions using parametric, two-dimensional, 
time-dependent hydrodynamic simulations. These models are such that
the linear stability properties are well understood, allowing the
exploration of well-differentiated regions of parameter 
space (Figure~\ref{f:growth_timescales}, Table~\ref{t:models}).
We have also introduced a spherical Fourier-Bessel decomposition
to characterize the properties of turbulence in the sub-sonic
region of the flow, to extract signatures of the interplay
of SASI and convection. Our main findings are as follows:
\newline

\noindent
1. -- The behavior of SASI-dominated models is characterized by the
      interplay of shock sloshings and the formation of large-scale,
      high-entropy structures. These bubbles are seeded by the SASI
      during shock expansions. Regular sloshing of the shock requires that
      these bubbles have a short lifetime and/or small entropy enhancements.
      Regular destruction of high-entropy structures by lateral flows
      is characteristic of non-exploding models (Figure~\ref{f:shock_entropy_sasi}).
      \newline

\noindent
2. -- Models that explode with SASI dominance are able to form large-scale,
      high-entropy bubbles that survive for a time longer than a 
      characteristic shock oscillation cycle (Figure~\ref{f:shock_entropy_sasi}). Neutrino heating and
      the inverse turbulent cascade in 2D ensure that these bubbles
      continue to grow if left undisturbed. Explosion results from the buoyancy
      of the bubble overcoming the drag force of the upstream flow \citep{thompson00}.
      \newline

\noindent
3. -- Convection-dominated models generate similar large-scale
      entropy structures by consolidating smaller-scale bubbles arising from
      buoyant activity. Sloshing of the shock occurs whenever large bubbles are
      destroyed or displaced, just as in SASI-dominated models, but without
      a dominant periodicity. The transition to explosion also involves the 
      formation and growth of a sufficiently
      large bubble (Figure~\ref{f:shock_entropy_conv}), as has been documented
      previously \citep{dolence2013,couch2012}. 
	\newline

\noindent 
4. -- Initial perturbations with a large amplitude do not alter the 
      qualitative way in which SASI-dominated models explode in two dimensions. 
      The difference in explosion time can be significant, however, and the
      time to runaway is not a monotonic function of the 
      perturbation amplitude (Figure~\ref{f:shock_entropy_pert}).
	\newline

\noindent
5. -- The time-averaged flow in convection-dominated, non-exploding models	
      adjusts itself to a state in which the convection parameter computed
      from the mean flow (eq.~[\ref{eq:chi_mean_flow}]) lies below
      the critical value for convective instability (Table~\ref{t:models}).
      This phenomenon is obscured when an average value of $\chi$ is computed
      from the instantaneous flow (Figure~\ref{f:brunt_comparison}).
	\newline

\noindent
6. -- The spherical Fourier-Bessel power in the subsonic, weakly-stratified
      region is dominated by
      the largest spatial scales (Figure~\ref{f:sfb_2d_mosaic}). The SASI 
      manifests itself as a characteristic
      even-odd pattern in the radial direction, and enhanced power at temporal frequencies
      corresponding to the advection time. Convection generates a smoother component,
      with power concentrated primarily below the advection frequency. This
      behavior of the frequency domain is consistent with the results of
      \citet{burrows2012} and \citet{mueller2012}.
	\newline

\noindent
7. -- The slope of the angular spectrum is consistent with 
      an inverse turbulent cascade at large $\ell$. Convection-dominated
      models yield angular spectra that resemble those of \citet{hanke2012}, 
      \citet{couch2012}, and \citet{dolence2013},
      while SASI-dominated models show a peak at $\ell=2$.
      The radial spectrum shows a scaling $n^{-2}$ at large $n$, which
      could be associated with Rayleigh-Taylor turbulence (e.g., \citealt{ciaraldi2009}).
      \newline

\noindent
8. -- The general results obtained with the parametric models persist when 
      the analysis is repeated on the  general relativistic, radiation-hydrodynamic simulations
      of \citet{mueller2012}. In particular, the behavior of the entropy when
      approaching explosion, and the value of the convection parameter and
      spectral slopes of the time-averaged flow are in good agreement with the
      corresponding parametric models.
      \newline

\noindent
9. -- The equality between advection and heating times in the gain region at $t=0$ is a good
      indicator of the onset of non-oscillatory instability in one-dimensional numerical
      simulations of parametric models (Appendix~\ref{s:L0}), in agreement with the
      numerical results of \citet{F12}. The fact that this
      equality occurs for heating rates such that the linear eigenmodes are still
      oscillatory (Figure~\ref{f:growth_timescales}), however, means that the onset of
      purely growing expansion is a non-linear
      effect (growth time shorter than the oscillation period). 
      Initial equality between the advection time in the gain and cooling layers
      is a good indicator of $\ell=0$ instability in some regions of the space of
      parametric models, but not in others, particularly when nuclear dissociation is included
      (Figure~\ref{f:growth_timescales}). When the recombination energy from alpha
      particles is not accounted for, the onset of $\ell=0$ instability does not necessarily
      lead to an explosion (Appendix~\ref{s:L0}), thus the instability thresholds do not
      equal explosion criteria for the parametric setup.
      \newline

Our results show that despite the non-linearity of the flow,
clear signatures of the operation of the SASI and convection can be obtained.
In particular, the parameter $\chi$ (equation \ref{eq:convection_parameter}) 
-- evaluated at the time where the shock stalls and
before hydrodynamic instabilities set in -- is a good predictor of whether the system
will be SASI- or convection-dominated on its way to explosion.

Despite the different explosion paths obtained when SASI or
convection dominate the dynamics at early times, it is not clear
that the resulting explosion properties are very different once
the process has started. Our results indicate that in both cases,
the formation of at least one large-scale, high-entropy bubble is a necessary
condition to achieve explosion in two-dimensions. It may be that
this degeneracy is triggered by the inverse turbulent cascade
inherent in axisymmetric models. 

The absence of this inverse cascade in 3D causes the flow to develop more
small-scale structure than in 2D (e.g., \citealt{hanke2012}). Nonetheless, the tendency
of bubbles to merge into bigger structures will persist, as this is
an intrinsic property of the Rayleigh-Taylor instability \citep{sharp1984}. In fact,
several 3D hydrodynamic studies have observed that prior to explosion,
a large-scale asymmetry (often $\ell=1$) develops in a non-oscillatory
way \citep{iwakami08,couch2012,dolence2013,hanke2013}. The difference lies
in the fact that the SASI provides seeds for large-scale entropy fluctuations
\emph{independent of dimension}, so it can speed up the formation of
a large- and hot enough bubble to achieve explosion. Verifying whether this picture
is robust requires numerical experiments in 3D.

Even though we have found clear evidence for high entropy bubbles
playing a key role in the interplay between SASI and convection
and in the onset of explosion, there are many questions that
remain to be answered. First, the evolution of the $f_V$ diagnostic
suggests that transition to explosion in a multidimensional environment
involves a fraction of the gain region volume achieving a certain entropy
or positive energy. What is that volume or mass fraction, and what are the
required entropy or energy values as a function of the dominant
system parameters? 

Second, our characterization of large bubble dynamics in the gain region
is limited. Processes such as seeding of bubbles by shock displacements,
survivability of bubbles due to neutrino heating, buoyancy, and the
turbulent cascade, disruption by lateral SASI flows in the linear phase
or low-entropy plumes in the non-linear phase, and feedback of these
bubbles on SASI modes deserve further study. Preliminary steps in this
direction have already been taken (e.g., \citealt{guilet09a,couch2012,dolence2013,murphy2012}),
though much more work remains if a quantitative understanding -- in the
form of a predictive explosion criterion -- is to be attained.

It is interesting to compare the critical heating rates for 
explosion in our parametric models and those from lightbulb setups with
a full EOS and a time-dependent mass accretion rate 
(e.g., \citealt{nordhaus10a,hanke2012,couch2012}). In the
former, explosion occurs above (but close to) the $\ell=0$ instability 
threshold in both SASI- and convection-dominated models, with only 
a $\sim 10\%$ difference between 1D and 2D (Fig.~\ref{f:growth_timescales}). 
In contrast, the latter models
are such that non-spherical instabilities make a larger difference
($\sim 20\%$) relative to the 1D case, with explosion occurring for
heating rates below the $\ell=0$ oscillatory instability. Note however
that both classes of models neglect the (negative) feedback to the
heating rate due to the drop in accretion luminosity when the shock
expands, thus the numbers obtained from these models should be treated with caution.
The search for a robust and predictive explosion criterion valid
for both SASI- and convection-dominated models is a worthwhile pursuit,
though outside the scope of the present paper.

The modification of our results by the introduction of a third
spatial dimension will be addressed in future work.

\section*{Acknowledgements}

We thank Jeremiah Murphy, Sean Couch, Christian Ott, J\'er\^ome Guilet, Adam Burrows, Josh Dolence, 
Yudai Suwa, Kei Kotake, Ernazar Abdikamalov, and Annop Wongwathanarat for stimulating discussions.
The authors thank the Institute for Nuclear Theory at the
University of Washington for its hospitality, and the US Department of Energy for
partial support during the completion of this work.
The anonymous referee provided helpful comments that improved
the presentation of the paper.
RF is supported by NSF grants AST-0807444, AST-1206097, and the University
of California Office of the President.
BM and HJ are supported by the Deutsche Forschungsgemeinschaft through the
Transregional Collaborative Research Center SBF-TR7 ``Gravitational Wave Astronomy"
and the Cluster of Excellence EXC 153 ``Origin and Structure of the Universe".
TF is supported by the grant ANR-10-BLAN-0503 funding the SN2NS project.
The software used in this work  was in part developed by 
the DOE NNSA-ASC OASCR Flash Center at the University of Chicago.
Parametric models were evolved at the IAS \emph{Aurora} cluster, while
the Garching models were evolved on the 
IBM p690 of the Computer Center Garching (RZG), on the Curie supercomputer
of the Grand \'Equipement National de Calcul Intensif (GENCI) under PRACE
grant RA0796, on the Cray XE6 and the NEC SX-8 at the HLRS
in Stuttgart (within project SuperN), and on the JUROPA systems at
the John von Neumann Institute for Computing (NIC) in J\"ulich.

\appendix

\section{On the Stability of the $\ell=0$ SASI mode}
\label{s:L0}

Here we compare the predictions from timescale ratio diagnostics 
with the actual eigenfrequencies of the $\ell=0$ mode in the parametric system.

\begin{figure}
\includegraphics*[width=\columnwidth]{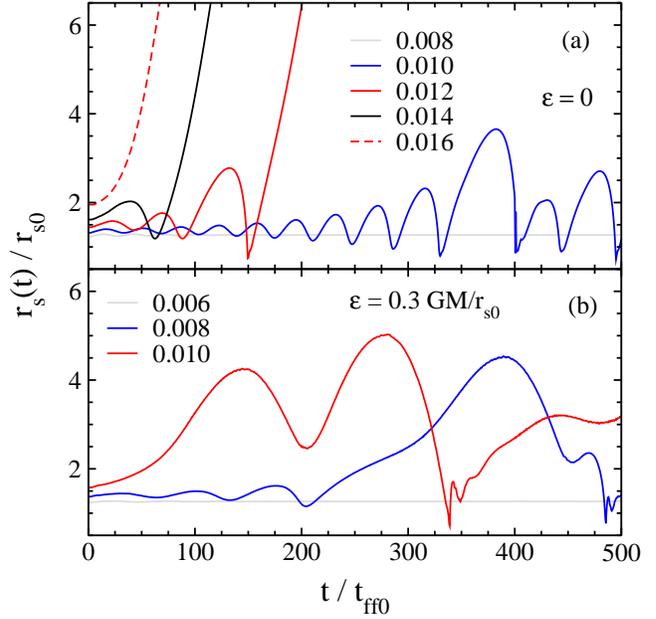}
\caption{Shock radius as a function of time for 1D models without dissociation 
(panel a) and with dissociation (panel b). Curves are labeled by the value
of the heating parameter $B$; compare with Figure~\ref{f:growth_timescales}. Note
that non-oscillatory expansion sets in when $B$ is such that $t_{\rm adv-g}>t_{\rm heat-g}$.
The shock expansion saturates in all models with dissociation, and in the model
with $B=0.01$ and $\varepsilon=0$.}
\label{f:shock_L0_paper}
\end{figure}

Figure~\ref{f:shock_L0_paper} shows the shock radius as a function of time for one dimensional
(1D) versions of the e0 and e3 sequences shown in Table~\ref{t:models}. By comparing with
Figure~\ref{f:growth_timescales}, one can see that the oscillatory radial stability thresholds are
well captured at this resolution. The initial value of the ratio of advection time in the gain region
to advection time in the cooling region is a good indicator of oscillatory radial stability
for the $\varepsilon=0$ sequence, but not so much when nuclear dissociation is introduced.

The initial ratio of advection to heating times in the gain region is a good predictor of non-oscillatory
expansion in the 1D models, in agreement with the numerical results of \citet{F12}.
Note however that for both sequences, this point lies at a lower heating rate
than the bifurcation of the perturbative $\ell=0$ growth rate (Fig.~\ref{f:growth_timescales}). Therefore,
this runaway expansion is a non-linear effect, likely arising from the fact that the growth
time is shorter than the oscillation period (by more than a factor of two in the e0 model when
the ratio of advection to heating timescales is unity).

Note also that in contrast to the models of \citet{F12}, radial instability does not 
always lead to runaway expansion. This is clear from the model with $\varepsilon=0$ and $B=0.01$,
which saturates. Also, all the models with nuclear dissociation saturate. \citet{FT09b} showed
that this effect is due to the artificial assumption of constant nuclear dissociation at the 
shock. Including the recombination energy of alpha particles as the shock
expands (which decreases the effective dissociation rate), leads to a runaway as soon
as instability sets in.

\section{Spherical Fourier-Bessel Decomposition in between Concentric Shells}
\label{s:sfb_appendix}

\begin{figure*}
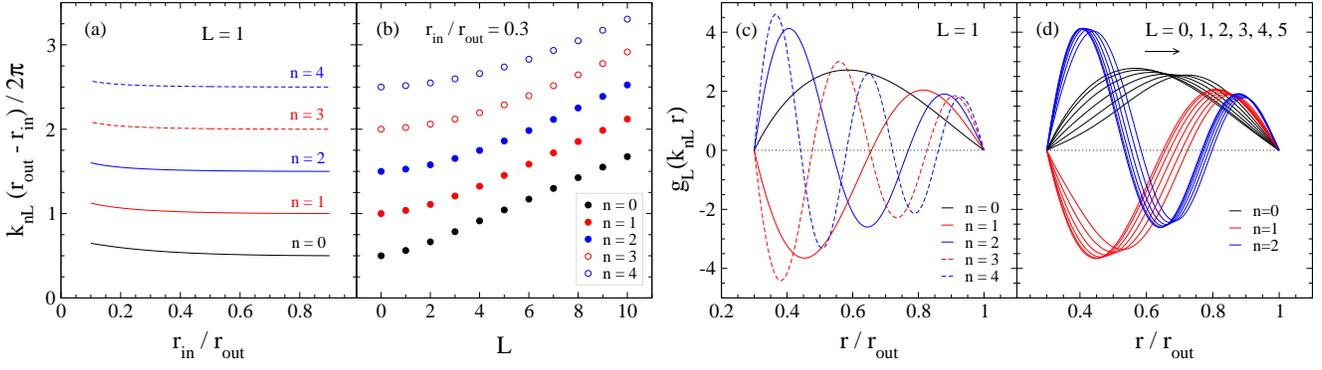

\includegraphics*[width=0.49\textwidth]{fB1a.eps}
\includegraphics*[width=0.49\textwidth]{fB1b.eps}
\caption{Properties of the radial eigenfunctions for Dirichlet boundary conditions. Panel
(a) shows wave numbers corresponding to the fundamental and first four harmonics as
a function of the domain size, for $\ell=1$. Panel (b) shows the corresponding wave numbers
when varying $\ell$, for fixed $r_{\rm in}/r_{\rm out} = 0.3$.
Panel (c) shows normalized eigenfunctions as a function of radius, for
the fundamental and the first four harmonics of $\ell=1$ and $r_{\rm in}/r_{\rm out}=0.3$. Panel
(d) shows the effect of varying $\ell$ from 0 to 5 on the fundamental and first two overtones,
for the same domain size as panel (c). Higher $\ell$ values shift curves to the right.}
\label{f:zeroes_eigenfunctions_dirichlet}
\end{figure*}

In spherical polar coordinates, the general solution to the Helmholtz equation\footnote{Since 
the Laplacian operator is Hermitian, its eigenfunctions -- solutions to the Helmholtz
equation -- form a complete orthogonal basis in the Hilbert space $L^2$ \citep{arfken05}.} 
is a superposition of functions of the form (e.g., \citealt{jackson})
\begin{equation}
\label{eq:helmholtz_solution}
\left[a_{\ell,m}j_\ell(kr) + b_{\ell,m}y_\ell(kr) \right]\,Y_{\ell}^m(\theta,\phi),
\end{equation}
where $j_\ell$ and $y_\ell$ are the spherical Bessel functions, $Y_\ell^m$ are
the Laplace spherical harmonics, and $\{a_{\ell,m},b_{\ell,m}\}$ are constant coefficients.
The wavenumber $k$ and the coefficients are determined once boundary conditions
for the problem are imposed at the radial domain boundaries $r_{\rm in}$ 
and $r_{\rm out}$.

\subsection{Dirichlet Boundary Conditions}

Requiring that the eigenfunctions vanish at the radial boundaries for all $\{\ell,m\}$
yields the system of equations
\begin{equation}
\label{eq:Dirichlet_condition}
\left[\begin{array}{cc} j_\ell(k\,r_{\rm in}) & y_\ell(k\,r_{\rm in})\\ 
                        j_\ell(k\,r_{\rm out}) & y_\ell(k\,r_{\rm out})\end{array} \right] 
\left( \begin{array}{c} a_{\ell,m}\\ b_{\ell,m}\end{array}\right) = 0.
\end{equation}
Non-trivial solutions are obtained by setting the determinant of the matrix of coefficients
to zero. This condition then defines a discrete set of radial wavenumbers:
\begin{eqnarray}
\label{eq:Dirichlet_wavenumbers}
j_\ell(k_{n\ell}\,r_{\rm in})\,y_\ell(k_{n\ell}\,r_{\rm out}) 
- j_\ell(k_{n\ell}\,r_{\rm out})\,y_\ell(k_{n\ell}\,r_{\rm in}) = 0\\
(n=0,1,2,...)\nonumber
\end{eqnarray}
where $n$ labels the roots in increasing magnitude.

Figure~\ref{f:zeroes_eigenfunctions_dirichlet}a shows the first five solutions for 
$\ell = 1$, as a function of the ratio of boundary radii $r_{\rm in}/r_{\rm out}$. 
We adopt the convention of labeling the smallest wavenumber by $n=0$, since the eigenfunction
has no nodes. For low $\ell$, the relation
\begin{equation}
\label{eq:wave_number_approx}
k_{n\ell}\simeq \frac{\pi}{(r_{\rm out}-r_{\rm in})}(n+1)\qquad n=0,1,2...
\end{equation}
holds approximately, becoming better for $r_{\rm in}/r_{\rm out}\to 1$.
Increasing the angular degree increases the value of the wave number relative to 
equation~(\ref{eq:wave_number_approx}), as shown in 
Figure~\ref{f:zeroes_eigenfunctions_dirichlet}b.

\begin{figure}
\begin{overpic}[width=\columnwidth]{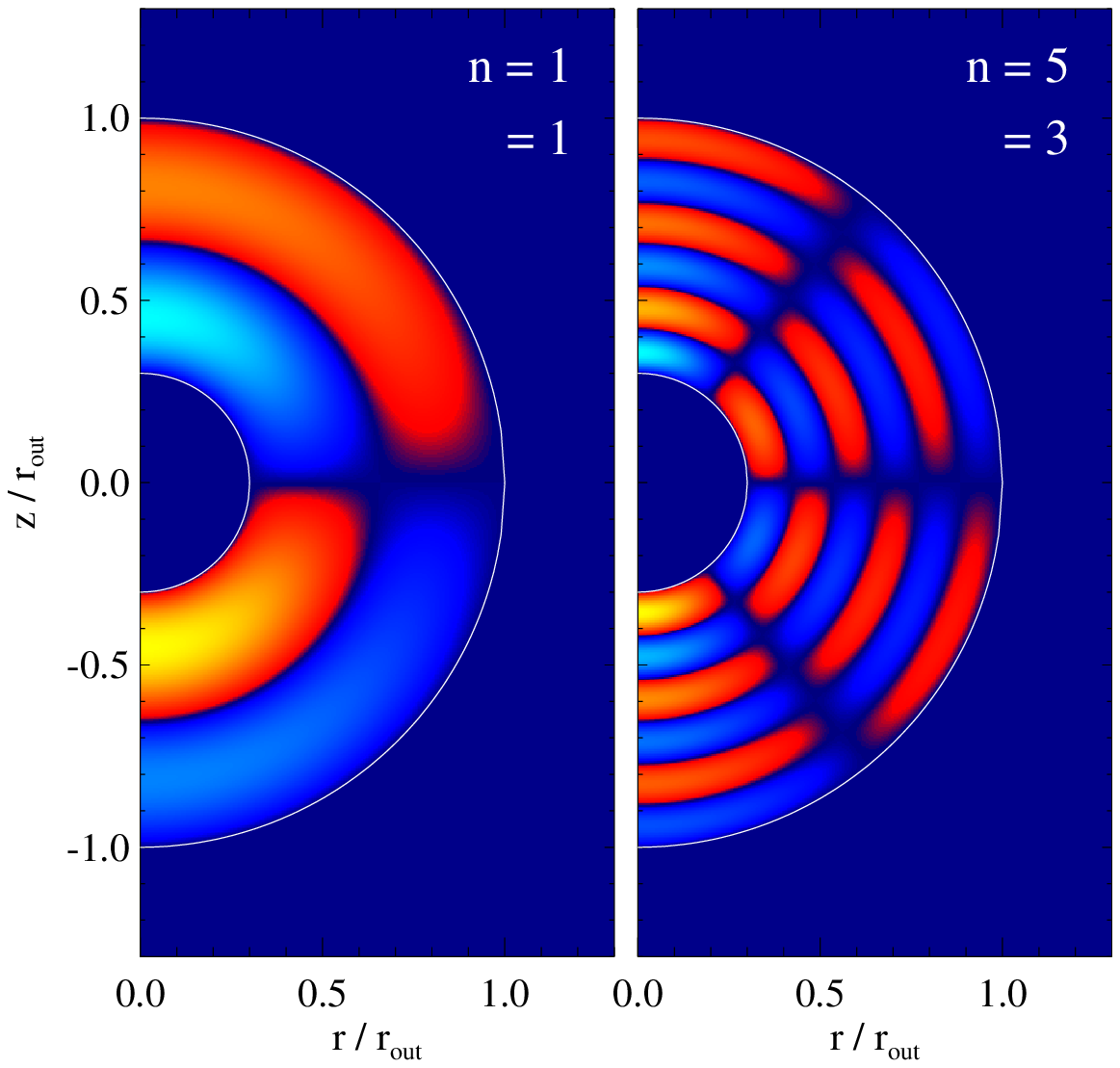}
\put(41,83){{\color{white} \Large $\ell$}}
\put(85,83){{\color{white} \Large $\ell$}}
\end{overpic}
\caption{Examples of spherical Fourier-Bessel basis functions in two dimensions $(r,\theta)$, for
a ratio of radii $r_{\rm in}/r_{\rm out} = 0.3$. Parameters are $\left\{n=1,\ell=1\right\}$ (left) and
$\left\{n=5,\ell=3\right\}$ (right). The basis functions vanish at the radial boundaries.}
\label{f:sample_eigenfunctions}
\end{figure}

Equation~(\ref{eq:Dirichlet_condition}) also determines the ratio of coefficients
\begin{equation}
\label{eq:Dirichlet_coefficients}
\frac{b_{n\ell m}}{a_{n\ell m}} = -\frac{j_\ell(k_{n\ell}\,r_{\rm in})}{y_\ell(k_{n\ell}\,r_{\rm in})} 
 			          = -\frac{j_\ell(k_{n\ell}\,r_{\rm out})}{y_\ell(k_{n\ell}\,r_{\rm out})}.
\end{equation}
Note that $n$ has been added as an index to the coefficients. The radial eigenfunctions $g_\ell$ are then
\begin{eqnarray}
\label{eq:first_formulation}
g_{n\ell}(r) & = &         N^{-1/2}_{n\ell}\left[y_\ell(k_{n\ell}\,r_{\rm out})j_\ell(k_{n\ell}\,r)\right.\nonumber\\ 
		&&\left.  - j_\ell(k_{n\ell}\,r_{\rm out})y_\ell(k_{n\ell}\,r)\right]\\
             & = & \tilde{N}^{-1/2}_{n\ell}\left[y_\ell(k_{n\ell}\,r_{\rm in})j_\ell(k_{n\ell}\,r)\right.\nonumber\\ 
		&&\left.  -  j_\ell(k_{n\ell}\,r_{\rm in})y_\ell(k_{n\ell}\,r)\right],
\end{eqnarray}
where the two formulations differ only by a global (real) phase. 
The normalization constant is found from the orthogonality condition 
(Lommel integral). Combining two solutions of the spherical 
Bessel differential equation, integrating over the radial domain, applying the boundary
conditions, and using L'H\^opital's rule yields
\begin{eqnarray}
\label{eq:normalization_condition_dirichlet}
\int_{r_{\rm in}}^{r_{\rm out}} g_\ell(k_{n\ell}\,r)g_\ell(k_{m\ell}\,r)\,r^2\totd r & = &
\frac{\delta_{nm}}{2}\left\{r_{\rm out}^3 \left[g^\prime_\ell(k_{n\ell}\,r_{\rm out})\right]^2\right.\nonumber\\
&&\left.- r_{\rm in}^3 \left[g^\prime_\ell(k_{n\ell}\,r_{\rm in})\right]^2\right\},
\end{eqnarray}
where $\delta_{nm}$ is the Kronecker symbol and primes denote derivative respect to the
argument. For the first formulation (eq.~\ref{eq:first_formulation}), choosing
\begin{eqnarray}
N_{n\ell} && = \frac{1}{2}\left\{r_{\rm out}^3 
\left[y_\ell(k_{n\ell}\,r_{\rm out})j^\prime_\ell(k_{n\ell}\,r_{\rm out})\right.\right.\nonumber\\ 
&&\left.  - j_\ell(k_{n\ell}\,r_{\rm out})y^\prime_\ell(k_{n\ell}\,r_{\rm out})\right]^2\nonumber\\
&& -r_{\rm in}^3 \left[y_\ell(k_{n\ell}\,r_{\rm out})j^\prime_\ell(k_{n\ell}\,r_{\rm in})\right. \nonumber\\
&&\left.\left.    - j_\ell(k_{n\ell}\,r_{\rm out})y^\prime_\ell(k_{n\ell}\,r_{\rm in})\right]^2\right\}
\end{eqnarray}
makes the eigenfunctions orthonormal.
Figure~\ref{f:sample_eigenfunctions} shows two examples of the resulting normalized eigenfunctions 
in a two dimensional, axisymmetric space. 

\begin{figure*}
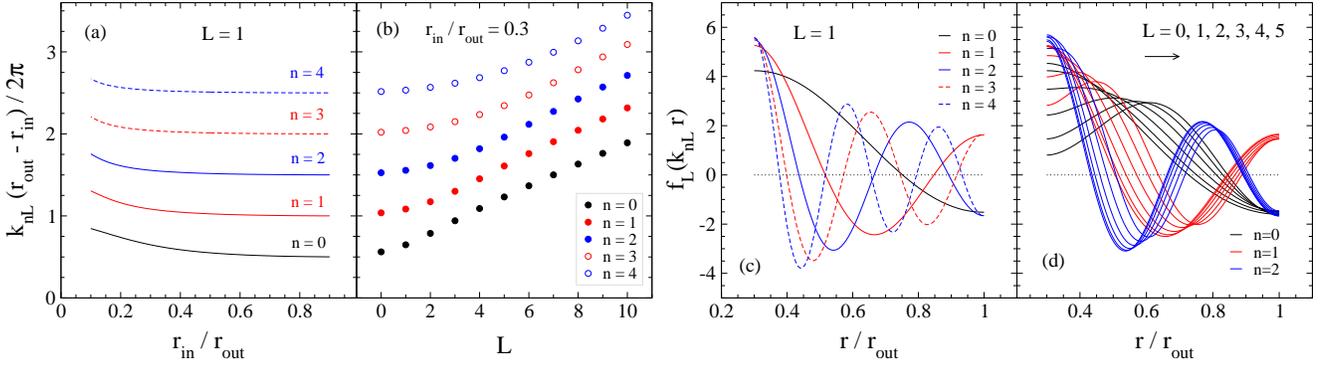

\includegraphics*[width=0.49\textwidth]{fB3a.eps}
\includegraphics*[width=0.49\textwidth]{fB3b.eps}
\caption{Same as Figure~\ref{f:zeroes_eigenfunctions_dirichlet}, but for Neumann boundary
conditions.}
\label{f:zeroes_eigenfunctions_neumann}
\end{figure*}

In three dimensions, the expansion of an arbitrary function
$f(r,\theta\,\phi)$ 
with Dirichlet boundary conditions in the radial interval $[r_{\rm in},r_{\rm out}]$ 
can be written as
\begin{equation}
\label{eq:expansion_functional_form}
f(r,\theta,\phi) = \sum_{n,\ell,m} f_{n\ell m} g_\ell(k_{n\ell}\,r) Y_\ell^m (\theta,\phi),
\end{equation}
with coefficients given by
\begin{equation}
f_{n\ell m} = \int r^2\totd r\,\totd\Omega\, g_\ell(k_{n\ell}\,r) Y_\ell^{m*}\,F(r,\theta,\phi),
\end{equation}
where the star denotes complex conjugation. The corresponding Parseval identity is
\begin{equation}
\int |F|\, \totd^3 x = \sum_{n,\ell,m} |a_{n\ell m}|^2
\end{equation}
yielding a three-dimensional spatial power spectrum:
\begin{equation}
\label{eq:PSD_3D}
P_{n\ell m} = |a_{n\ell m}|^2.
\end{equation}

\subsection{Neumann Boundary Conditions}

Requiring that the radial derivative of the eigenfunctions vanish at
the boundaries yields the equation for the radial wave numbers
\begin{eqnarray}
\label{s:Neumann_wavenumbers}
j^\prime_\ell(k_n\,r_{\rm in})\,y^\prime_\ell(k_n\,r_{\rm out}) 
- j^\prime_\ell(k_n\,r_{\rm out})\,y^\prime_\ell(k_n\,r_{\rm in})
= 0\\
\quad (n=1,2,3 ...),\nonumber
\end{eqnarray}
where the primes again denote derivative respect to the argument.
The eigenfunctions are now
\begin{eqnarray}
\label{eq:eigenfunctions_neumann_form1}
f_\ell(k_n\,r)& = & M^{-1/2} \left[y^\prime_\ell(k_n\,r_{\rm in})j_\ell(k_n\,r)\right.\nonumber\\
             &&\left.  -  j^\prime_\ell(k_n\,r_{\rm in})y_\ell(k_n\,r)\right]\\
\label{eq:eigenfunctions_neumann_form2}
              & = & \tilde{M}^{1/2} \left[y^\prime_\ell(k_n\,r_{\rm out})j_\ell(k_n\,r)\right.\nonumber\\ 
	     &&  \left. - j^\prime_\ell(k_n\,r_{\rm out})y_\ell(k_n\,r)\right],
\end{eqnarray}
and the orthogonality condition reads
\begin{eqnarray}
\label{eq:normalization_condition_neumann}
\int_{r_{\rm in}}^{r_{\rm out}} f_\ell(k_n\,r)f_\ell(k_m\,r)\,r^2\totd r  = &&\nonumber\\ 
\frac{\delta_{nm}}{2}
\left\{r_{\rm out}^3 \left[1-\frac{\ell(\ell+1)}{(k_{n\ell}r_{\rm out})^2} \right]f^2_\ell(k_{n\ell}\,r_{\rm out})\right.\nonumber\\
\qquad\left.- r_{\rm in}^3 \left[1 - \frac{\ell(\ell+1)}{(k_{n\ell}r_{\rm in})^2} \right]f^2_\ell(k_{n\ell}\,r_{\rm in})  \right\}.
\end{eqnarray}
The normalization constant for equation~(\ref{eq:eigenfunctions_neumann_form1}) is
\begin{eqnarray}
M_{n\ell} & = & \frac{1}{2}\left\{
             r_{\rm out}^3\left[1 - \frac{\ell(\ell+1)}{(k_{n\ell}r_{\rm out})^2}\right]\right.\times\nonumber\\
         &&                 \left[y_\ell^\prime(k_{n\ell}r_{\rm out})\,j_\ell(k_{n\ell}r_{\rm out})
             -j^\prime_\ell(k_{n\ell}r_{\rm out})\,y_\ell(k_{n\ell}r_{\rm out})\right]^2\nonumber\\
          &&      -r_{\rm in}^3\left[1 - \frac{\ell(\ell+1)}{(k_{n\ell}r_{\rm in})^2}\right]
                     \left[y_\ell^\prime(k_{n\ell}r_{\rm out})\,j_\ell(k_{n\ell}r_{\rm in})\right.\nonumber\\
          &&\left.\left.\qquad\qquad\qquad\quad
             -j^\prime_\ell(k_{n\ell}r_{\rm out})\,y_\ell(k_{n\ell}r_{\rm in})\right]^2.
             \right\} 
\end{eqnarray}

The radial wave numbers and eigenfunctions for the first few harmonics and $\ell$ values are shown in 
Figure~\ref{f:zeroes_eigenfunctions_neumann}. The overall structure of the wave numbers is
very similar to the Dirichlet case, with slightly higher values for small ratio of radii and
large $\ell$. For fixed harmonic, the eigenfunctions
change their shape as $\ell$ is increased, in contrast to the Dirichlet case.

\bibliographystyle{mn2e}
\bibliography{ccsne,apj-jour}

\label{lastpage}
\end{document}